\documentclass[12pt]{article}
    \newcommand{\be}[1]{\begin{equation}\label{#1}}
    \newcommand{\ba}[1]{\begin{eqnarray}\label{#1}}
    \newcommand{\ep}[1]{\epsilon_{#1}}

    \newcommand{\pa}[1]{\left(#1\right)}
    \newcommand{\paq}[1]{\left[#1\right]}
    
    \newcommand{\av}[1]{\langle#1\rangle}
    \newcommand{\M}{{\rm M_{\rm P}}}
       \newcommand{\Mt}{{\rm \widetilde M_{\rm P}}}

    \def\ee{\end{equation}}
    \def\ea{\end{eqnarray}}

\usepackage{graphicx,latexsym}
\usepackage{graphicx,epsf}
\usepackage{color}
\usepackage{epsfig}
\usepackage{amsmath}
\usepackage[T1]{fontenc}
\usepackage[utf8]{inputenc}
\usepackage{authblk}
\begin{document}
%%%%%%%%%%%%%
\title{Quantum Cosmology and the Evolution of Inflationary Spectra}

\author[1,2]{Alexander Y. Kamenshchik\thanks{Alexander.Kamenshchik@bo.infn.it}}
\author[1]{Alessandro Tronconi\thanks{Alessandro.Tronconi@bo.infn.it}}
\author[1]{Giovanni Venturi\thanks{Giovanni.Venturi@bo.infn.it}}
\affil[1]{Dipartimento di Fisica e Astronomia and INFN, Via Irnerio 46,40126 Bologna,
Italy}
\affil[2]{L.D. Landau Institute for Theoretical Physics of the Russian
Academy of Sciences, Kosygin str. 2, 119334 Moscow, Russia}

\renewcommand\Authands{ and }
\maketitle
%%%%%%%%%%%%%

\begin{abstract}
We illustrate how it is possible to calculate the quantum gravitational effects on the spectra of primordial scalar/tensor perturbations starting from the canonical, Wheeler-De Witt, approach to quantum cosmology. The composite matter-gravity system is analysed through a Born-Oppenheimer approach in which gravitation is associated with the heavy degrees of freedom and matter (here represented by a scalar field) with the light ones. Once the independent degrees of freedom are identified the system is canonically quantised. The differential equation governing the dynamics of the primordial spectra with its quantum-gravitational corrections is then obtained and is applied to diverse inflationary evolutions. Finally, the analytical results are compared to observations through a Monte Carlo Markov Chain technique and an estimate of the free parameters of our approach  is finally presented
and the results obtained are compared with previous ones.
 \end{abstract}
%\date{}
%%%%%%%%%%%%%%%%%%
\section{Introduction}
The paradigm of inflation \cite{inflation}
has led to a beautiful connection between microscopic and macroscopic scales. This occurs since inflation acts as a ``magnifying glass'' insofar as microscopic quantum fluctuations at the beginning of
time, when the universe was very small, evolve into inhomogeneous structures \cite{Stewart:1993bc}. Thus the observed structure of the present-day universe is related to the very early time quantum  dynamics. As a consequence the former can be used to test the primordial dynamics and in particular the  possible effects of quantum gravity at early times corresponding to a very small universe. The reason for this is that because of the huge value of the Planck mass quantum gravity effects are otherwise suppressed (of course one can also hope to observe quantum gravitational effects in the presence of very strong gravitational fields, for example in the proximity of black holes).\\
Composite systems which involve two mass (or time) scales such as molecules are amenable to treatment by a Born-Oppenheimer approach \cite{BO}. For molecules this is possible because of the different  nuclear and electron masses, this allows one to suitably factorise the wave-function of the composite system leading, in a first approximation, to
a separate description of the motion of the nuclei and the electrons. In particular it is found that the former are influenced by the mean hamiltonian of the latter and the latter  (electrons) follow the former adiabatically (in the quantum mechanical sense).  Similarly for  the matter gravity system as a consequence  of the fact that gravity  is characterised by the Planck mass, which is much greater than the usual matter mass, the heavy degrees of freedom are associated with gravitation and the light ones with matter \cite{BO-cosm}. As a consequence, to lowest order, gravitation will be driven by the main matter Hamiltonian and matter will follow gravity adiabatically.
As mentioned above we shall quantise the composite system, by this we mean that we shall perform the canonical quantisation of Einstein gravity and matter  leading to the Wheeler-DeWitt (WDW) equation \cite{DeWitt}. This  is what we mean by quantum gravity and is quite distinct to the introduction of so-called trans-Planckian effects (loosely referred to as quantum gravity) through ad hoc modifications of the dispersion relation \cite{Martin} and/or the initial conditions \cite{inicond}.  Further the equations we shall obtain after the BO  decomposition will be exact, in the sense that they also include non-adiabatic effects.
The above  approach has been previously illustrated in a mini-superspace model with the aim of studying the semiclassical emergence of time \cite{BO-cosm}, which is otherwise absent in the quantum system. Conditions were found for the usual (unitary) time evolution of quantum matter (Schwinger-Tomonaga or Schr\"odinger) to emerge, essentially these are that non-adiabatic transitions (fluctuations) be negligible or that the universe be sufficiently far from the Planck scale.
In a series of papers \cite{K} we have generalized the approach to non-homogeneous cosmology in order to obtain corrections to the usual power spectrum of cosmological fluctuations produced during inflation. These corrections, which essentially amount to the inclusion of the effect of the non-adiabatic transitions,  affect the infrared part of the spectrum and lead to an amplification or a suppression depending on the background evolution. More interestingly they depend on the wavenumber $k$ and scale as $k^{-3}$, in both the scalar and the tensor sectors, when background evolution is close to de Sitter. That non-adiabatic effects  affect the infrared part of the spectrum, which is associated with large scales, is not surprising, since it is this part of the spectrum which exits the horizon in the early stages of inflation and is exposed to high energy and curvature effects for a longer time.

The latest Planck mission results \cite{cmb} provide the most accurate constraints available currently to inflationary dynamics \cite{inflation}. So far the slow roll (SR) mechanism has been confirmed to be a paradigm capable of reproducing the observed spectrum of cosmological fluctuations and the correct tensor to scalar ratio \cite{Stewart:1993bc}. Since the inflationary period is the cosmological era describing the transition from the quantum gravitational scale down to the hot big bang scale, it may, somewhere, exhibit related  peculiar features which could be associated with quantum gravity effects. Quite interestingly a loss of power, with respect to the expected flatness for the spectrum of cosmological perturbations, can be extrapolated from the data at large scales \cite{powerloss1}.  Since, as mentioned above, it is for such scales that quantum gravity effects  due to non-adiabaticity may appear, this has motivated us to estimate such effects. Unfortunately, such a feature (evident already in the WMAP results) exhibits large errors due to cosmic variance. Nonetheless we feel that it is worth comparing our detailed analytical predictions for the quantum gravity effects with Planck data  through a Monte Carlo Markov Chain (MCMC) based method.\\
The paper is organized as follows. In  section 2 the basic equations are reviewed, the canonical quantization method and the subsequent BO decomposition are illustrated. In the section 3 we calculate the master equation governing the dynamics of the two point function of the quantum fluctuations when the quantum gravitational effects are taken into account and the vacuum prescription for these fluctuations is briefly discussed. In  section 4 we review the basic relations for de Sitter, power law and slow-roll (SR) inflation and the quantum corrections to the primordial spectra are explicitly calculated for these three distinct cases. In section 5 we illustrate how our analytical predictions are compared to observations and we comment our results. Finally in section 6 we draw the conclusions.

%%%%%%%%%%%%%%%%%%%%%
\section{Basic equations}
The inflaton-gravity system is described by the following action
\be{fullaction}
S=\int d\eta d^3x\sqrt{-g}\paq{-\frac{\M^2}{2}R+\frac{1}{2}\partial_\mu\phi\partial^\mu\phi-V(\phi)}
\ee
where $\M=\pa{8\pi G}^{-1/2}$ is the reduced Planck mass. The above action can be decomposed into a homogeneous part plus fluctuations around it. The fluctuations of the metric $\delta g_{\mu\nu}(\vec x,\eta)$ are defined by
\be{metricsc}
g_{\mu\nu}=g_{\mu\nu}^{(0)}+\delta g_{\mu\nu}
\ee  
where $g_{\mu\nu}^{(0)}=\rm{diag}\paq{a(\eta)^2\pa{1,-1,-1,-1}}$ and $\eta$ is the conformal time. Only the scalar and the tensor fluctuations ``survive'' the inflationary expansion: $\delta g=\delta g^{(S)}+\delta g^{(T)}$. The scalar fluctuations of the metric can be defined as follows
\be{metricsc2}
\delta g_{\mu\nu}=a(\eta)^2\left(
\begin{matrix} 
2 A(\vec x,\eta) & -\partial_i B(\vec x,\eta)\\
\\
-\partial_i B(\vec x,\eta) & 2\delta_{ij}\psi(\vec x,\eta)-D_{ij}E(\vec x,\eta)
\end{matrix}\right)
\ee
with $D_{ij}\equiv \partial_i\partial_j-\frac{1}{3}\delta_{ij}\nabla^2$. These four degrees of freedom (d.o.f.) mix with the inflation fluctuation $\delta \phi(\vec x,\eta)$, defined by $\phi(\vec x,\eta)\equiv \phi_0 (\eta)+\delta \phi(\vec x,\eta)$. The scalar perturbations, defined in (\ref{metricsc2}), are gauge dependent. One can either rewrite them in terms of just two Bardeen's potentials or fix the gauge and set two of them to zero. Finally, on using the equations of motion, the scalar sector can be collectively described by a single field $v(\vec x,\eta)$ which, in the uniform curvature gauge, is given by $v(\vec x,\eta)=a(\eta)\delta \phi(\vec x,\eta)$. Its Fourier transform, $v_k$, can then be decomposed into two parts: $v_{1,k}\equiv {\rm Re}\pa{v_k}$ and $v_{2,k}\equiv {\rm Im}\pa{v_k}$. \\ 
The tensor fluctuations are gauge invariant perturbations of the metric and are defined by
\be{metrich}
ds^2=a(\eta)^2\paq{d\eta^2-\pa{\delta_{ij}+h_{ij}}dx^idx^j}
\ee
with $\partial^i h_{ij}=\delta^{ij}h_{ij}=0$. For each direction of propagation of the perturbation $k^i$,
the above conditions on $h_{ij}$, with the requirement $g_{\mu\nu}=g_{\nu\mu}$, give seven independent constraint equations for the components of the tensor perturbations, leading to only two remaining polarization physical degrees of freedom $h^{(+)}$ and $h^{(\times)}$. Then, on defining $v_{1,k}^{(\lambda)}\equiv \frac{a \M}{\sqrt{2}}\rm{Re}\pa{h_k}$ and $v_{2,k}^{(\lambda)}\equiv \frac{a \M}{\sqrt{2}}\rm{Im}\pa{h_k}$, one can describe the tensor perturbations in a manner similar to the scalar perturbations.\\
In what follows we shall illustrate in detail a point which is often glossed over: namely the fact that on working in a flat 3-space and considering both homogeneous and inhomogeneous quantities one must introduce an unspecified length $L$. Indeed the effective action of the homogeneous inflaton-gravity system plus the inhomogeneous perturbations finally is \cite{MukMald}
%The effective action describing the dynamics of the homogeneous inflaton-gravity system plus the perturbations finally is \cite{MukMald}
\ba{act}
S&=&\int d\eta\left\{L^3\paq{-\frac{\Mt^2}{2}a'^2+\frac{a^2}{2}\pa{\phi_0'^2-2V(\phi_{0})a^2}}\right.\nonumber\\
&+&\left.\frac{1}{2}\sum_{i=1,2}\sum_{k\neq 0}^\infty\paq{v_{i,k}'(\eta)^2+\pa{-k^2+\frac{z''}{z}}v_{i,k}(\eta)^2}\right.\nonumber\\
&+&\left.\frac{1}{2}\sum_{i=1,2}\sum_{\lambda=+,\times}\sum_{k\neq 0}^\infty\paq{\pa{\frac{v_{i,k}^{(\lambda)}}{d\eta}}^2+\pa{ -k^2 +\frac{a''}{a}}\pa{v_{i,k}^{(\lambda)}}^2 }
\right\}
\ea
where $\Mt=\sqrt{6}\M$, $z\equiv \phi_0'/H$, $H=a'/a^2$ is the Hubble parameter and $L^3\equiv \int d^3x$. Let us note that the action for the perturbations has been conveniently simplified by means of the homogeneous dynamics.\\
The interval $ds$ has dimension of a length $l$ and one generally may either take $[a]=l$ and $[dx]=[d\eta]=l^0$ or $[a]=l^0$ and $[dx]=[d\eta]=l$. Correspondingly one then has $[L]=l^0$ or $[L]=l$. One can eliminate the factor $L^3$ by replacing $a\rightarrow a/L$, $\eta\rightarrow \eta L$, $v\rightarrow \sqrt{L} v$ and $k\rightarrow k/L$. Such a redefinition is equivalent to setting $L=1$ in the above action (\ref{act}) (then implicitly assuming the convention $[a(\eta)]=l$ and $[dx]=[d\eta]=l^0$) and then proceeding with its quantization. Such a choice, although limited to the homogeneous part, has been previously illustrated \cite{FVV}. 
Henceforth we shall use this latter simplifying choice. Only at the end, in order to compare our results with observations, we shall restore all quantities to their original definition and the dependence on $L$ will become explicit. 
Let us finally note that the fact that $L$ is infinite does not create a problem. As usual, the transition from the Fourier integral w.r.t. the wave number to the Fourier series eliminates the corresponding divergence.  \\
%Let us note that the dependence on $L$ can always be restored  by returning on setting back $a$, $\eta$, $v$ and $k$ to their original definition.\\
Once the total action for the matter-gravity system is cast into the form (\ref{act}), all the dynamical quantities (fields) are expressed through an infinite ``tower'' of homogeneous variables $v_{i,k}$. Such an effective description has a simplifying role in the quantization procedure, which we shall illustrate in detail in the next section.  

%%%%%%%%%%%%%%%%%%%%%%%

\subsection{Canonical Quantization}
The dynamics of each d.o.f. describing the perturbations, is formally analogous to that of a homogeneous scalar field with a time dependent mass. In order to illustrate the quantization procedure and the subsequent Born-Oppenheimer decomposition in detail, without losing generality,  we single out the homogenous part and one real scalar field for the perturbations in (\ref{act}):
\ba{actS}
S&=&\int d\eta\left\{\paq{-\frac{\Mt^2}{2}a'^2+\frac{a^2}{2}\pa{\phi_0'^2-2V(\phi_{0})a^2}}\right.\nonumber\\
&+&\left.\frac{1}{2}\sum_{k\neq 0}^\infty\paq{v_{k}'(\eta)^2-\omega_k^2v_{k}(\eta)^2}\right\}\equiv\int d\eta \mathcal{L}_{tot}
\ea
where $\omega_k^2=k^2+m^2(\eta)$ is time dependent and $L$ has been set equal to $1$. Let us note that $m^2(\eta)$ depends on the homogeneous quantities $a(\eta)$, $\phi_0(\eta)$ and their derivatives. The action describing the evolution of the cosmological perturbations, is derived by substituting the homogenous, leading order, solutions into the perturbed Lagrangian. Such a derivation does not affect the quantization of the perturbations but may have consequences on the quantization of the homogeneous d.o.f.. Let us remember  that in obtaining the reduced action (\ref{actS}) we have at most kept terms to quadratic order in the field and metric perturbations ($v_k$). Therefore, since quantum fluctuations around $z''/z$ occur already multiplied by small field perturbations, we shall just retain for it its classical homogeneous value. Thus our choice is to consider $m^2(\eta)$ as a generic function of time and consequently specify it at the end of the quantization procedure.\\

One can rewrite the above action in terms of an arbitrary time parameter $\tau$ with $N(\tau)d\tau=a(\eta)d\eta$, where $N(\tau)$ is the lapse function. The action (\ref{actS}) then becomes 
\ba{actN}
S&=&\int d\tau \frac{N}{a}\left\{\paq{-\frac{\Mt^2}{2}\frac{a^2\dot{a}^2}{N^2}+\frac{a^4}{2}\pa{\frac{\dot{\phi_0}^2}{N^2}-2V(\phi_{0})}}\right.\nonumber\\
&+&\left.\frac{1}{2}\sum_{k\neq 0}^\infty\paq{\frac{a^2\dot{v}_{k}(\eta)^2}{N^2}-\omega_k^2v_{k}(\eta)^2}\right\}\equiv\int d\tau \tilde{\mathcal{L}}_{tot}
\ea
where the dot indicates the derivative w.r.t. $\tau$. The lapse function plays the role of a Lagrange multiplier in the action. The variation of the action w.r.t $N$ leads to the following equation of motion
\be{hamcon}
0=\frac{\delta \tilde{\mathcal{L}}_{tot}}{\delta N}=\frac{\Mt^2}{2}\frac{a\dot a^2}{N^2}-\frac{a^3\dot\phi_0^2}{2N^2}-a^3 V-\sum_{k\neq 0}^\infty\paq{\frac{a\dot v_k^2}{2N^2}+\frac{\omega_k^2v_k^2}{2a}}
\ee  
having the form of a constraint equation.
The system Hamiltonian is 
\be{ham}
\mathcal{H}=-\frac{N\pi_a^2}{2a\Mt^2}+\frac{N\pi_\phi^2}{2a^3}+a^3NV+\sum_{k\neq 0}^{\infty}\paq{\frac{N\pi_k^2}{2a}+\frac{N\omega_k^2}{2a}v_k^2}
\ee
where
\be{momenta}
\pi_N=0\,,\;\pi_a=-\frac{\Mt^2a\dot a}{N}\,,\;\pi_\phi=\frac{a^3\dot \phi_0}{N}\,,\;\pi_k=\frac{a\dot v_k}{N}.
\ee
and is proportional to the above constraint (\ref{hamcon}):
\be{hamcon2}
0=\frac{\delta \tilde{\mathcal{L}}_{tot}}{\delta N}=\frac{\mathcal H}{N}
\ee
which is then called ``Hamiltonian constraint''. It is a very particular energy conservation constraint which equates the system's total energy to zero. At the quantum level, when the degrees of freedom are canonically quantized, it plays the role of a time independent Schroedinger equation.

The canonical quantization of the action (\ref{act}) leads to the following Wheeler-De Witt (WDW) equation \cite{DeWitt} for the wave function of the universe (matter plus gravity)
\begin{eqnarray}
&&\left\{\frac{1}{2\Mt ^2}\frac{\partial^2}{\partial a^2}-\frac{1}{2a^2}\frac{\partial^2}{\partial \phi_0^2}+Va^4\right.\nonumber\\
&&\left.+\sum_{k\neq 0}^{\infty}\paq{-\frac{1}{2}\frac{\partial^2}{\partial v_k^2}+\frac{\omega_k^2}{2}v_k^2}\right\}\Psi\pa{a,\phi_0,\{v_k\}}=0\label{WDW0}.
\end{eqnarray}
Let us note that the time dependent mass in $\omega_k^2$ is $m^{2}(\eta)=-\frac{z''}{z}$ for each mode of the scalar perturbation and $m^{2}(\eta)=-\frac{a''}{a}$ for each mode of the tensor perturbation, where $z(\eta)$, $a(\eta)$ are classical expressions.
%%%%%%%%%%%%%%%%%%%%%%%%%%%%
\subsection{Born-Oppeneheimer decomposition}
Eq. (\ref{WDW0}) can be written in the compact form 
\ba{WDW}
&&\paq{\frac{1}{2\Mt^2}\frac{\partial^2}{\partial a^2}+\hat H_0^{(M)}+\sum_k \hat H_k^{(M)}}\Psi\pa{a,\phi_0,\{v_k\}}\nonumber\\
&&\equiv\paq{\frac{1}{2\Mt^2}\frac{\partial^2}{\partial a^2}+\hat H^{(M)}}\Psi\pa{a,\phi_0,\{v_k\}}=0.
\ea 
where 
\be{HM0}
\hat H_0^{(M)}=-\frac{1}{2a^2}\frac{\partial^2}{\partial \phi_0^2}+Va^4,
\ee
\be{HMk}
\hat H_k^{(M)}=-\frac{1}{2}\frac{\partial^2}{\partial v_k^2}+\frac{\omega_k^2}{2}v_k^2
\ee
and is formally similar to a time independent Schroedinger equation, except for the sign in front of the kinetic term for the scale factor.
Finding the general solution of the WDW equation, even when the perturbations are set to zero, is a very complicated task due to the interaction between matter and gravity.\\
A set of approximate solutions can be found within a BO approach. The BO approximation was originally introduced in order to simplify the Schroedinger equation of complex atoms and molecules \cite{BO}.

It consists in factorising the wave function of the Universe into a product
\be{BOdec}
\Psi\pa{a,\phi_0,\{v_k\}}=\psi(a)\chi\pa{a,\phi_0,\{v_k\}}
\ee
where $\psi(a)$ is the wave function for the homogeneous gravitational sector and $\chi\pa{a,\phi_0,\{v_k\}}$ is that for matter (homogeneous plus perturbations). A similar decomposition for atoms consists in factorising the atomic wave function $\Psi_A(r,R)$ into a nuclear wave function $\psi_N(R)$ and the  electrons' wave functions $\chi_e(r,R)$, where $r$ and $R$ are the d.o.f. of electrons and nuclei respectively. The matter wave function in eq. (\ref{BOdec}) can be further factorized as:
\be{MATfac}
\chi\pa{a,\phi_0,\{v_k\}}=\chi_0\pa{a,\phi_0}\prod_{k\neq 0}^\infty \chi_k\pa{\eta,v_k}=\prod_{k=0}^{\infty}\chi_k.
\ee
Let us note that the wave function of each mode $v_k$ depends parametrically on the conformal time $\eta$ and, in the semiclassical limit, the evolution of the scale factor $a=a(\eta)$ fixes $\eta$ as a function of $a$. The above factorization leads to the following set of partial differential equations, which are equivalent to the WDW equation:
\be{graveq}
\paq{\frac{1}{2\Mt^2}\frac{\partial^2}{\partial a^2}+\av{\hat H^{(M)}}}\tilde \psi=
-\frac{1}{2\Mt^2}\langle\frac{\partial^2}{\partial a^2}\rangle\tilde \psi
\ee
which is the equation for the gravitational wave function and
\ba{mateq}
&&\tilde\psi^*\tilde\psi\paq{\hat H^{(M)}-\av{\hat H^{(M)}}}\tilde \chi+\frac{1}{\Mt^2}\pa{\tilde\psi^*\frac{\partial}{\partial a}\tilde \psi}\frac{\partial}{\partial a}\tilde \chi\nonumber\\
&&=\frac{1}{2\Mt^2}\tilde\psi^*\tilde\psi\paq{\av{\frac{\partial^2}{\partial a^2}}-\frac{\partial^2}{\partial a^2}}\tilde \chi
\ea
which is the equation for matter, where
\be{gder}
\psi=e^{-i\int^{a} \mathcal{A} da'}\tilde\psi,\;  \chi=e^{i\int^{a} \mathcal{A} da'}\tilde\chi, \; \mathcal{A}=-i\langle\chi|\frac{\partial}{\partial a}|\chi\rangle
\ee
with $v_0=\phi_0$, $\av{\hat O}=\langle\tilde \chi|\hat O|\tilde\chi\rangle$ and each mode is individually normalized by $\langle \chi_{k}|\chi_{k}\rangle=\int d v_k\chi_k^*\chi_k=1$. 
The r.h.s. of eqs. (\ref{graveq}) and (\ref{mateq}) are associated with non adiabatic quantum effects. They are generally neglected in the leading order to the BO approximation.\\
On multiplying both sides by $\hat P_k=\prod_{j\neq k} \langle \tilde \chi_k|$  eq. (\ref{mateq}) can be split into a set of equations, each governing the dynamics of a single mode $k$ of the matter field. One is then led to 
\ba{mateqk}
&&\tilde\psi^*\tilde\psi\paq{\hat H_k^{(M)}-\langle \tilde \chi_k|\hat H_{k}^{(M)}|\tilde \chi_k\rangle}\tilde \chi_k+\frac{1}{\Mt^2}\pa{\tilde\psi^*\frac{\partial \tilde \psi}{\partial a}}\nonumber\\
&&\times \frac{\partial \tilde \chi_k}{\partial a}=\frac{1}{2\Mt^2}\tilde\psi^*\tilde\psi\paq{\langle \tilde \chi_k|\frac{\partial^2}{\partial a^2}|\tilde \chi_k\rangle-\frac{\partial^2}{\partial a^2}}\tilde \chi_k.
\ea
We may now perform the semiclassical limit for the gravitational wave function $\psi(a)$ by setting 
\be{semicl}
\tilde\psi(a)\sim(\Mt^{2}a')^{1/2}\exp\pa{-i\int^{a}\Mt^{2}a' da}
\ee
obtaining the Friedmann equation
\be{geqsc}
-\frac{\Mt^2}{2}a'^2+\sum_k\av{\hat H_k^{(M)}}=0
\ee
for Eq. (\ref{graveq}), to the leading order. In such a way the BO decomposition of the wave function of the universe is uniquely determined and $a$ and $\eta$  are related.\\
Now, on defining $|\chi_{k}\rangle_{s}\equiv e^{-i\int^\eta\langle \tilde \chi_k|\hat H_{k}^{(M)}|\tilde \chi_k\rangle d\eta'}|\tilde \chi_{k}\rangle$, Eq. (\ref{mateqk}) becomes
\ba{evoeq}
\!\!\!\!\!\!\!&&\!\!\!\!\!\!\!i\partial_{\eta}|\chi_{k}\rangle_{s}-\hat H_{k}^{(M)} |\chi_{k}\rangle_{s}=\frac{\exp\paq{{i\int^\eta \langle \tilde \chi_k|\hat H_{k}^{(M)}|\tilde \chi_k\rangle d\eta'}}}{2\Mt^2}\nonumber\\
\!\!\!\!\!\!\!&&\!\!\!\!\!\!\!\times\paq{\partial_a^2-\frac{a''}{(a')^{2}}\partial_{a}-\langle\tilde \chi_{k}|\pa{\partial_a^2-\frac{a''}{(a')^{2}}\partial_{a}}|\tilde \chi_{k}\rangle} |\tilde\chi_k\rangle\nonumber\\
\!\!\!\!\!\!\!&&\!\!\!\!\!\!\!\equiv\epsilon \paq{\hat \Omega_{k}-\av{\hat \Omega_{k}}_{s}}|\chi_{k}\rangle_{s}
\ea
where $\av{\hat O}_{s}\equiv \phantom|_{s}\langle \chi_{k} |\hat O|\chi_{k}\rangle_{s}$ and $\epsilon\equiv\frac{1}{2\Mt^{2}}$. In Eq. (\ref{evoeq}) we have retained all terms, in order to consistently include contributions to $\mathcal{O}\pa{\Mt^{-2}}$ (different expansions have been previously examined and compared for the homogeneous case \cite{BO-unit}). 
The operator $\hat \Omega_{k}$ has the following form:
\be{Omega}
\hat \Omega_{k}=\frac{1}{a'^{2}}\frac{d^{2}}{d\eta^{2}}+\paq{2i\frac{\av{\hat H_{k}^{(M)}}_{s}}{a'^{2}}-2\frac{a''}{a'^{3}}}\frac{d}{d\eta}.
\ee
The operator on the r.h.s. of Eq. (\ref{evoeq}) has a nonlinear structure, since it depends on $\chi_{s}$ and $\chi_{s}^{*}$ through multiplicative factors of the form $\av{\hat O}_{s}$. 
We immediately note that ,in the absence of the r.h.s., Eq. (\ref{evoeq}) becomes the usual matter evolution equation (Schr\"odinger or Schwinger-Tomonaga). The terms on the r.h.s. describe the non-adiabatic effects of quantum gravitational origin.\\
%%%%%%%%%%%%%%%%%%%%%%%%%%%%%%%%
\section{Two point function}
We are interested in the observable features of the spectrum of the scalar/tensor fluctuations generated during inflation. Such features can be extracted from the two-point function 
\be{two}
p(\eta)\equiv \!\!\!\!\phantom{A}_{s}\langle 0|\hat v^{2}|0\rangle_{s}=\av{\hat v^{2}}_{0}
\ee
at late times (for the modes well outside the horizon). In (\ref{two}) the vacuum state $|0\rangle_{s}$ satisfies the full equation (\ref{evoeq}) and, according to standard prescriptions, reduces to the Bunch-Davies (BD) vacuum \cite{BD} in the short wavelength regime (more general assumptions may be considered as well). Let us note that $p(\eta)$ also depends on $k$ but, in order to keep notation compact, we decided to omit any explicit reference on it.

%%%%%%%%%%%%%%%%%%%%%%%%
\subsection{Unperturbed dynamics}
Before tackling the problem of evaluating the evolution of $p(\eta)$ by taking into account the full dynamics given by (\ref{evoeq}), in this section we shall briefly review the basic formalism for the unperturbed dynamics.\\
For each $k$ mode, on neglecting the quantum gravitational effects,  Eq. (\ref{evoeq}) takes the form of a time dependent Schr\"odinger equation for a harmonic oscillator with time dependent frequency
\be{H0}
\hat H_{k}^{(M)}=\frac{\hat \pi_{k}^{2}}{2}+\frac{\omega_{k}^{2}}{2}\hat v_{k}^{2}
\ee
where $\omega_k=\omega_k(\eta)$.
The subscript $k$ and the label $(M)$ will henceforth be omitted. The following consideration will be valid for both scalar and tensor perturbations.\\
At the classical level, $v$ and $\pi$ satisfy the Hamiltonian equations leading to the homogeneous classical Klein-Gordon equation (equation of a harmonic oscillator with a time dependent frequency) :
\be{cleq}
v''+\omega^{2} v=0.
\ee
At the quantum level, the solutions of the time dependent Schroedinger equation can be found by introducing a linear invariant operator $\hat I$, satisfying the differential equation
\be{inveq}
i \frac{d}{d\eta}\hat I+\paq{\hat I,\hat H}=0
\ee
and building up a complete set of states from the invariant vacuum state $|{\rm vac}\rangle$, defined by $\hat I|{\rm vac}\rangle=0$, and then iteratively applying $\hat I^{\dagger}$ to the vacuum. A linear invariant satisfying (\ref{inveq}) is given by
\be{Idef}
I=i\paq{\varphi^{*}\hat \pi-\pa{\varphi^{*}}'\hat v}
\ee
where $\varphi^{*}$ satisfies the classical equation of motion (\ref{cleq}). The commutator satisfies $\paq{\hat I,\hat I^{\dagger}}=1$, provided the wronskian condition
\be{wrc}
i \paq{\varphi^{*}\varphi'-\pa{\varphi^{*}}'\varphi}=1
\ee
holds. Then, in the coordinate representation, the properly normalised invariant vacuum is
\be{vacI}
\langle v|{\rm vac}\rangle=\paq{\frac{1}{2\pi\pa{\varphi^{*}\varphi}}}^{1/4}\exp\paq{\frac{i}{2}\frac{\pa{\varphi^{*}}'}{\varphi^{*}}v^{2}}
\ee
and a suitable phase is needed in order for $|{\rm vac}\rangle$ to satisfy the Schroedinger equation. One easily finds
\be{vacls}
|0\rangle_s=\exp\paq{{-\frac{i}{4}\int^\eta\frac{d\eta'}{\varphi^*\varphi}}}|{\rm vac}\rangle.
\ee
Let us note that the wronskian condition, (\ref{wrc}), does not fix the invariant vacuum in a unique way. In general, different linearly independent combinations of solutions of eq. (\ref{cleq}), satisfying the wronskian condition, are allowed. The BD prescription is only one of the possible choices. Consequently the expression (\ref{vacI}) is a more general vacuum state satisfying the unperturbed quantum dynamics.\\
%%%%%%%%%%%%%%%%%%%%%%%%%%%%%
The linear invariants may be alternatively defined in terms of the so-called Pinney variable. In particular $\hat I$ can be written as:
\be{InvP}
\hat I=\frac{e^{i\Theta}}{\sqrt{2}}\paq{\pa{\frac{1}{\rho}-i\rho'}\hat v+i\rho\hat \pi}
\ee
where $\rho$ is the Pinney variable, a real function satisfying the following non linear differential equation (the so-called Ermakov--Pinney equation \cite{Erm-Pin})
\be{pin}
\rho''+\omega^{2}\rho=\frac{1}{\rho^{3}}
\ee
with $\Theta=\int^{\eta} \frac{d\eta'}{\rho^{2}}$. In terms of $\rho$ the commutator $\paq{\hat I,\hat I^{\dagger}}=1$ is now trivially satisfied.
The Pinney variable is related to the solution $\varphi$ of the classical field equation (\ref{cleq}) by 
\be{pvar}
\rho=\sqrt{2\varphi^{*}\varphi}
\ee
hence it is proportional to its modulus.
In the coordinate representation, the properly normalised vacuum, expressed in terms of the Pinney, variable is
\be{vacBD}
\langle v|0\rangle_s=\frac{1}{\pa{\pi\rho^2}^{1/4}}\exp\paq{-\frac{i}{2}\int^\eta \frac{d\eta'}{\rho^2}-\frac{v^2}{2}\pa{\frac{1}{\rho^2}-i\frac{\rho'}{\rho}}}.
\ee
Let us finally note that the two point function is given by
\be{pun}
p(\eta)=\varphi^*\varphi=\frac{\rho^2}{2}.
\ee
%%%%%%%%%%%%%%%%%%%%%%%%%%%%%
\subsection{Perturbed evolution}
When quantum gravitational effects are taken into account, one must solve the integro-differential equation (\ref{evoeq}), which is an extremely difficult task.\\
Instead of trying to solve (\ref{evoeq}) and then calculating the power spectrum, one can find the differential equation for the spectrum $p$, by iteratively differentiating the two-point function and using the canonical commutation relations.\\
On taking $|\chi_{k}\rangle_{s}=|0\rangle_{s}$ in Eq. (\ref{evoeq}) (we are omitting the subscript $k$) one obtains the evolution equation for the vacuum
\ba{evo0} 
\!\!\!0&=&\!\!\!i\frac{d}{d\eta}|0\rangle_{s}-\hat H|0\rangle_{s}-\left[\left(2i\av{\hat H}_{0} g\pa{\eta}+g'\pa{\eta}\right)\right.\nonumber\\
\!\!\!&\times&\!\!\!\left.\pa{\frac{d}{d\eta}-\av{\frac{d}{d\eta}}_{0}}+g(\eta)\pa{\frac{d^{2}}{d\eta^{2}}-\av{\frac{d^{2}}{d\eta^{2}}}_{0}}\right]|0\rangle_{s}
\ea
with $\av{\hat O}_{0}\equiv \!\!\!\!\phantom{A}_{s}\langle 0|\hat O|0\rangle_{s}$ and $g(\eta)=\frac{1}{2\Mt^{2}a'^{2}}$.
The evolution of the two-point function can be now calculated by differentiating (\ref{two}) w.r.t. $\eta$ and using (\ref{evo0}). The first derivative of $p$ w.r.t. the conformal time is
\be{p1}
i\frac{dp}{d\eta}=\av{\left[\hat v^{2},\hat H\right]}_{0}-\av{\hat v^{2}}_{0}F(\eta)+G_{\hat v^{2}}(\eta)
\ee
where
\be{Fdef}
F(\eta)=\pa{2i g\av{\hat H}_{0}+g'}\av{\partial_{\eta}}_{0}+g\av{\partial_{\eta}^{2}}_{0}-c.c.\;,
\ee
\be{Gdef}
G_{\hat v^{2}}(\eta)=\pa{2i g\av{\hat H}_{0}+g'}\av{\hat v^{2}\partial_{\eta}}_{0}+g\av{\hat v^{2}\partial_{\eta}^{2}}_{0}-c.c..
\ee
Let us note that $g$ is a real function and $F$ and $G_{\hat v^{2}}$ are then purely imaginary functions of $\eta$ by construction. The subscript $\hat v^{2}$ in (\ref{Gdef}) indicates that the function $G$ depends on $\eta$ and on the operator $\hat v^{2}$. 
The commutator in the expression (\ref{p1}) is $\left[\hat v^{2},\hat H\right]=i\left\{\hat v,\hat \pi\right\}$.
In a more compact form Eq. (\ref{p1}) can then be written as
\be{p1com}
\frac{d\av{\hat v^{2}}_{0}}{d\eta}=\av{\left\{\hat v,\hat \pi\right\}}_{0}-iR(\hat v^{2})
\ee
where $R$ contains the quantum gravitational effects and is defined as $R(\hat O)=-\av{\hat O}_{0}F(\eta)+G_{\hat O}(\eta)$.
The above expression can be differentiated once more w.r.t. $\eta$ and takes the following form
\be{p2}
\frac{d^{2}\av{\hat v^{2}}_{0}}{d\eta^{2}}=\frac{d\av{\left\{\hat v,\hat \pi\right\}}_{0}}{d\eta}-i\frac{d R(\hat v^{2})}{d\eta}.
\ee
and, in analogy with (\ref{p1}):
\be{p2term}
\frac{d\av{\left\{\hat v,\hat \pi\right\}}_{0}}{d\eta}=-i\av{\left[\left\{\hat v,\hat \pi\right\},\hat H\right]}_{0}-iR\left(\left\{\hat v,\hat \pi\right\}\right).
\ee
The commutator in the expression above becomes $\left[\left\{\hat v,\hat \pi\right\},\hat H\right]=2i\pa{\hat \pi^{2}-\omega^{2}\hat v^{2}}$ and (\ref{p2}) can be then rewritten as
\be{p2com}
\frac{d^{2}\av{\hat v^{2}}_{0}}{d\eta^{2}}=2\pa{\av{\hat \pi^{2}}_{0}-\omega^{2}\av{\hat v^{2}}_{0}}-iR\left(\left\{\hat v,\hat \pi\right\}\right)-i\frac{d R(\hat v^{2})}{d\eta}.
\ee
On then calculating  the derivative of  Eq. (\ref{p2com}) we finally obtain:
\ba{p3}
\frac{d^{3}\av{\hat v^{2}}_{0}}{d\eta^{3}}\!\!\!&=&\!\!\!\frac{d\av{\hat \pi^{2}}_{0}}{d\eta}-4\omega\omega'\av{\hat v^{2}}_{0}-2\omega^{2}\frac{d\av{\hat v^{2}}_{0}}{d\eta}\nonumber\\
\!\!\!&-&\!\!\!i\frac{dR\left(\left\{\hat v,\hat \pi\right\}\right)}{d\eta}-i\frac{d^{2} R(\hat v^{2})}{d\eta^{2}},
\ea	
where
\be{com3}
\frac{d\av{\hat \pi^{2}}_{0}}{d\eta}+iR(\hat \pi^{2})=-i\av{\left[\hat \pi^{2},\hat H\right]}_{0}
=i\omega^{2}\av{\left[\hat v^{2},\hat H\right]}_{0}
\ee
and 
\be{com4}
\av{\left[\hat v^{2},\hat H\right]}_{0}=i\frac{d\av{\hat v^{2}}_{0}}{d\eta}-R(\hat v^{2}).
\ee
Equation (\ref{p3}) finally becomes
\ba{p33}
0\!\!\!&=&\!\!\!\frac{d^{3}\av{\hat v^{2}}_{0}}{d\eta^{3}}+4\omega^{2}\frac{d\av{\hat v^{2}}_{0}}{d\eta}+2\pa{\omega^{2}}'\av{\hat v^{2}}_{0}+2iR(\hat \pi^{2})\nonumber\\
\!\!\!&+&\!\!\!2i\omega^{2}R(\hat v^{2})+i\frac{dR\left(\left\{\hat v,\hat \pi\right\}\right)}{d\eta}+i\frac{d^{2} R(\hat v^{2})}{d\eta^{2}}.
\ea
Let us note that eq. (\ref{p33}) is exact (no simplifications have been done to obtain eq. (\ref{p33}) starting from (\ref{evo0})). Further eq. (\ref{p33}) has been obtained without using  any peculiar property of the vacuum state and is also valid for any state satisfying the modified Schroedinger equation (\ref{evoeq}).

A perturbative approach is needed in order to solve eq. (\ref{p33}). To the first order in $\Mt^{-2}$, one can then evaluate the quantum gravitational corrections on the unperturbed vacuum (\ref{vacBD}) and then identify $\rho\rightarrow \sqrt{2p}$.
The differential master equation governing the evolution of the two point function is finally
\be{qfx}
\frac{d^{3}p}{d\eta^{3}}+4\omega^{2}\frac{dp}{d\eta}+2\frac{d\omega^{2}}{d\eta}p+\Delta_p=0
\ee
with
\be{qfxP}
\Delta_p=-\frac{1}{\M^{2}}\left[\frac{d^{3}}{d\eta^{3}}\frac{h}{4 a'^{2}}-\frac{d^{2}}{d\eta^{2}}\frac{p'\pa{h+2}}{4 pa'^{2}}-\frac{d}{d\eta}\frac{h^2+4p'^2}{8a'^{2}p^{2}}+\frac{\omega\omega'h}{a'^{2}}\right]
\ee
where 
\be{defhp}
h\equiv p'^{2}+4\omega^{2}p^{2}-1.
\ee
The above equation is valid to the first order in $\Mt^{-2}$ and, in the $\Mt\rightarrow \infty$ limit, it must reproduce the standard evolution of the two point function, which is known to satisfy the second order differential equation 
\be{pequn}
\frac{d^2p}{d\eta^2}-\frac{1}{2p}\pa{\frac{dp}{d\eta}}^2+2\omega^2 p-\frac{1}{2p}=0.
\ee
as can be easily derived from (\ref{pin}), given the relation (\ref{pun}). Differentiation of eq. (\ref{pequn}) leads to the third order equation (\ref{qfx}), without quantum gravitational effects.\\
The above master equation can be used for the evolution of the vacuum (and not of a generic quantum state). Let us observe that (\ref{qfx}) is a third order differential equation for $p$ and also contains unphysical solutions, which do not satisfy the unperturbed eq. (\ref{pequn}) in the $\Mt\rightarrow \infty$ limit.

%%%%%%%%%%%%%%%%%%%%%%%%%%%%%%%%%%
\subsection{The vacuum prescription}
In the short wavelength limit $-k\eta\gg 1$, the classical equation (\ref{cleq}) admits plane wave solutions of the form $v_{\pm}=\frac{1}{\sqrt{2k}}\exp\pa{\pm ik\eta}$ and an arbitrary linear combination provides a suitable initial state of the system. In particular, if one retains only positive frequency waves, correspondingly one has $p=\frac{1}{2k}$. The initial condition $p=\frac{1}{2k}$ corresponds to the so-called BD vacuum prescription. The BD vacuum state is the quantum state which coincides with the Hamiltonian vacuum, as initial condition. Let us note that, in the short wavelength limit, $h$ is zero for $p=1/2k$ and consequently, to the leading order, the quantum gravitational corrections calculated in our approach are $\Delta_p=0$.  \\
The general combination of plane wave solutions is
\be{genv}
v=\frac{1}{\sqrt{2k}}\pa{\alpha \exp\pa{ik\eta}+\beta \exp\pa{-ik\eta}}
\ee
corresponding to
\be{genp}
p=\frac{1}{2k}\paq{|\alpha|^2+|\beta|^2+2{\rm Re}\pa{\alpha \beta^*\exp\pa{2ik\eta}}}.
\ee
The integration constants $\alpha$ and $\beta$ are complex numbers, constrained by the wronskian condition (\ref{wrc}) which leads to 
\be{wrc0}
|\alpha|^2-|\beta|^2=1;
\ee
the BD vacuum simply corresponds to $|\beta|=0$. One may rewrite the expression for $p$, given the condition (\ref{wrc0}), and find
\be{pwrc}
p=\frac{1}{2k}\paq{1+2|\beta|^2+2|\beta|\sqrt{1+|\beta|^2}\pa{\cos \delta\cos 2k\eta-\sin \delta\sin 2k\eta}}
\ee
where $\delta$ is the difference between the phases of $\alpha$ and $\beta$ respectively. Let us note that 2 real parameter ($\delta$ and $|\beta|$) enter the final expression (\ref{pwrc}), playing the role of the 2 integration constants of the second order differential equation (\ref{pequn}). Let us note that, only for $|\beta|=0$, $h=0$ and the quantum gravitational corrections are negligible in the short wavelength limit.\\ 
If one solves the third order differential equation (\ref{qfx}), even on neglecting the quantum corrections, 3 integrations constants are necessary for the general solution. However only a subset of these solutions is physical, i.e. satisfy eq. (\ref{pequn}), and one then expects some relation holds among the three integration constants. On solving the eq. (\ref{qfx}), in the short wavelength limit ($-k\eta\gg 1$) and $\Mt\rightarrow \infty$, one finds
\be{shortfullDS}
p\simeq\frac{1}{2k^{2}}\paq{c_{+}-c_{-}\cos\pa{2k\eta}+c_{0}\sin\pa{2k\eta}}.
\ee
then on comparing with (\ref{pwrc}) we have
\be{condc1}
\frac{c_+}{k}=1+2|\beta|^2,\; \frac{c_-}{k}=-2|\beta|\sqrt{1+|\beta|^2}\cos\delta,\;\frac{c_0}{k}=-2|\beta|\sqrt{1+|\beta|^2}\sin\delta
\ee
or equivalently
\be{condc2}
c_+^2-c_-^2-c_0^2=k^2,\;c_+>0.
\ee
The BD vacuum corresponds to $c_-=c_0=0$.\\
%%%%%%%%%%%%%%%%%%%%%%%%%%%%
\section{Applications}
In this section we apply our formalism to diverse inflationary backgrounds and calculate the quantum gravitational corrections to primordial spectra. In particular we study a pure de Sitter evolution, power law inflation and finally SR inflation. Our starting point is the equation 
\be{peqper}
\frac{d^2p}{d\eta^2}-\frac{1}{2p}\pa{\frac{dp}{d\eta}}^2+2\omega^2 p-\frac{1}{2p}=-\frac{1}{p}\int_{-\infty}^{\eta}d\eta' p\Delta_p
\ee
which is obtained by integrating (\ref{qfx}) and  imposing the BD initial conditions on $p$, i.e. $p(-\infty)=1/(2k)$, $p'(-\infty)=p''(-\infty)=0$.
%%%%%%%%%%%%%%%%%%%%%%%%%%%
\subsection{De Sitter evolution}

In order to illustrate the main effects of quantum gravity on the spectrum, starting from unperturbed  exact expressions, the de Sitter case is first discussed. Such a case can be obtained from realistic inflationary models in the limit $\dot H\rightarrow 0$, at least for $\Delta_p=0$.\\
When $H = \rm{const}$, one has $\omega=\sqrt{k^{2}-\frac{2}{\eta^{2}}}$ for both scalar and tensor perturbations (the equation for the scalar sector must be obtained by starting from a general background evolution and then  taking the $\dot H\rightarrow 0$ limit).\\
The BD solution of Eq. (\ref{pequn}) is 
\be{pDSun}
p=\frac{1+k^2\eta^2}{2k^3\eta^2}
\ee
leading to the following expression for $\Delta_p$:
\be{depDS}
\Delta_p=\frac{4H^{2}}{\Mt^{2}k^{4}\eta^{3}}=-\frac{4H^2}{k\Mt^2}p'
\ee
to the first order in $\Mt$. Then  eq. (\ref{peqper}) can be rewritten as
\be{peqper2}
\frac{d^2p}{d\eta^2}-\frac{1}{2p}\pa{\frac{dp}{d\eta}}^2+2\omega^2 p-\frac{1}{2p}=\frac{2H^2}{k\Mt^2p}\pa{p^2-p_\infty^{2}}
\ee
with $p_{\infty}=1/(2k)$. The latter equation can be recast in the form of the original, unperturbed equation (\ref{pequn}), by defining
\be{tildepds}
\tilde p=\frac{p}{\sqrt{1-\frac{4H^2}{\Mt^2k}p_{\infty}^2}}
\ee 
and
\be{tildeomegads}
\tilde\omega^2=\omega^2-\frac{H^2}{\Mt^2k}\equiv \tilde k^2-\frac{z''}{z}
\ee
with 
\be{deftildek}
\tilde k=k\sqrt{1-\frac{H^2}{\Mt^2k^3}}\equiv N_k k.
\ee
The general solution of
\be{peqper3}
\frac{d^2\tilde p}{d\eta^2}-\frac{1}{2\tilde p}\pa{\frac{d\tilde p}{d\eta}}^2+2\tilde \omega^2 \tilde p-\frac{1}{2\tilde p}=0
\ee
is known and is given by
\ba{fullsol}
\tilde p&=&\frac{1}{2\tilde k^{4}\eta^{2}}\left\{\sqrt{\tilde k^2+c_0^2+c_-^2}\pa{1+\tilde k^{2}\eta^{2}}+\cos\pa{2\tilde k\eta}\left[2c_{0}\tilde k\eta\right.\right.\nonumber\\
&&\left.\left.-c_{-}\pa{\tilde k^{2}\eta^{2}-1}\right]+\sin\pa{2\tilde k\eta}\left[c_{0}\pa{\tilde k^{2}\eta^{2}-1}+2c_{-}\tilde k\eta
\right]\phantom{A \over B}\!\!\!\!\!\!\!\right\}.
\ea
On setting the oscillatory contribution to zero ($c_{-}=c_{0}=0$), one finally finds the perturbed BD vacuum
\be{BDDS}
p=\frac{1+N_k^2k^{2}\eta^{2}}{2N_k^2 k^{3}\eta^{2}}.
\ee
In the long wavelength limit, one finds the observable features of the primordial spectra
\be{pDSlong}
p\stackrel{-k\eta\rightarrow0}{\longrightarrow}\frac{1}{2k^{3}\eta^{2}\pa{1-\frac{H^2}{\Mt^2k^3}}}
\ee
and, for $\frac{H^2}{\Mt^2k^3}\ll 1$, such spectra behave as
\be{DSlong2}
p\stackrel{-k\eta\rightarrow0}{\longrightarrow}\frac{1}{2k^{3}\eta^{2}}\pa{1+\frac{H^2}{\Mt^2k^3}}=p_0\pa{1+\frac{H^2}{\Mt^2k^3}}
\ee
i.e. quantum gravitational effects lead to a power enhancement w.r.t. the standard results in the spectrum for large scales.\\
Let us note that the length scale $L$, defined in the section 2, is hidden in the expression for the quantum gravitational corrections. On returning to the original physical quantities one has $p\rightarrow p/L$, $a\rightarrow L a$, $\eta\rightarrow \eta/L$ and $k\rightarrow L k$. The scale $L\equiv \bar k^{-1}$ would then appear in the result, as an effect of the initial volume integration of the homogeneous dynamics.

%%%%%%%%%%%%%%%%%%%%%%%%%%%%
\subsection{Power-Law}
Power-law inflation corresponds to the simplified case in which the Hubble parameter depends on time, yet still the equations of motions, for both the homogeneous part and the perturbations, can  be solved exactly. In this case the evolution of the scale factor is given by
\be{PLaeta}
a(\eta)=a_0\pa{\frac{\eta_0}{\eta}}^\frac{q}{q-1}.
\ee
where $q$ is a constant parameter, which is  related to the variation of $H$ by $q\equiv \pa{-\dot H/H^2}^{-1}$ and the de Sitter limit is recovered for $q\rightarrow \infty$.
The dynamics of the scalar and the tensor perturbations are governed by the same equation, which is given by (\ref{peqper}) with $\omega=\sqrt{k^{2}-\frac{2}{\eta^{2}}\frac{1-\frac{\epsilon}{2}}{(1-\epsilon)^2}}$. The BD vacuum can be now expressed in term of the Hankel functions:
\be{pPLun}
p=-\frac{\pi\eta}{4}\paq{H_\nu^{\pa{1}}(-k\eta)H_\nu^{\pa{2}}(-k\eta)}
\ee
with $\nu=\frac{3}{2}+\frac{1}{q-1}$. The observable features of the primordial spectra can be calculated by taking the long wavelength limit of (\ref{pPLun}) , finding
\be{plongPL0}
p\rightarrow-\frac{\eta}{4\pi}\Gamma(\nu)^2\pa{-\frac{k\eta}{2}}^{-2\nu}=\paq{2^{\frac{q+1}{q-1}}\frac{\Gamma(\nu)^2}{\pi}}\eta_0^{-\frac{2q}{q-1}}k^{-2\nu}\pa{\frac{a}{a_0}}^2.
\ee
Alternatively one may solve (\ref{pequn}) in the long wavelength regime.
In this case one simply observes that $\omega^2\rightarrow -a''/a$, $p\rightarrow C_0 a^2$ and the normalization, $C_0$, can be fixed ,by matching the long wavelength solution with the short wavelength prescription for the BD vacuum ($p\rightarrow 1/(2k)$) at the horizon crossing ($k=a_k H_k$), namely:
\be{matchPLun}
C_0 \frac{k^2}{H_k^2}=\frac{1}{2k}\Rightarrow C_0=\frac{1}{2ka_k^2}.
\ee
One  then obtains
\be{plongPL}
p\rightarrow\frac{1}{2k}\pa{\frac{a}{a_k}}^2=\frac{1}{2}\paq{\frac{q}{(q-1)}}^{\frac{2q}{q-1}}\eta_0^{-\frac{2q}{q-1}}k^{-2\nu}\pa{\frac{a}{a_0}}^2
\ee
On comparing the results (\ref{plongPL0}) and (\ref{plongPL}), we observe that the normalization constant $C_0$ ,obtained by the matching procedure, is very close to the exact normalization when $q$ is large (and they coincide in the $q\rightarrow \infty$ limit, i.e. for the de Sitter case).

One can also adopt the matching procedure to solve the perturbed equation (\ref{qfx}). We already observed that the quantum gravitational corrections are negligible to  leading order, in the short wavelength limit. Conversely they can be evaluated perturbatively and then the long wavelength limit taken. In such a limit we find that
\be{perPL}
p\Delta_p\rightarrow A_0 \pa{p^2}'
\ee
with
\be{A0} 
A_0=-\frac{2C_0k^2}{\Mt^2}\frac{\pa{q-1}\pa{2q+1}}{q\pa{q+1}} 
\ee
where $C_0$ is the normalization of $p$.\\
On neglecting $\Delta_p$ in the interval $\left]-\infty,\eta_k\right]$, one is then led to the following perturbed equation, valid in the long wavelength regime
\be{peqper2}
\frac{d^2p}{d\eta^2}-\frac{1}{2p}\pa{\frac{dp}{d\eta}}^2+2\omega^2 p-\frac{1}{2p}=-\frac{1}{p}\int_{\eta_k}^{\eta}d\eta' p\Delta_p
\ee
where $\eta_k$ is the conformal time at the horizon crossing $a_k H_k=k$. Let us note that (\ref{peqper2}) is now obtained by integrating (\ref{qfx}) and, on imposing the conditions $p(\eta_k)=1/(2k)$, $p'(\eta_k)=p''(\eta_k)=0$.\\
The integral on the r.h.s. of eq. (\ref{peqper2}) can be easily performed, in the long wavelength regime given (\ref{perPL}), and  takes the form:
\be{peqperPL}
\frac{d^2p}{d\eta^2}-\frac{1}{2p}\pa{\frac{dp}{d\eta}}^2+2\omega^2 p-\frac{1}{2p}+\frac{A_0}{2}p-\frac{A_0}{4k^2p}=0.
\ee
On defining $\tilde k=k\sqrt{1+\frac{A_0}{2k^2}}$ and $\tilde p=p/\pa{\sqrt{1+\frac{A_0}{2k^2}}}$, this latter equation can be cast in the form of the unperturbed equation , having  the following solution
\be{ptilPL}
\tilde p=\tilde C_0 a^2.
\ee
The normalization factor $\tilde C_0$ can here be fixed, by matching the short and the long wavelength solutions at the horizon crossing, i.e. when $a_{\tilde k}H_{\tilde k}=\tilde k$. One then finds
\be{matchPLtil}
\frac{1}{2 k\pa{\sqrt{1+\frac{A_0}{2k^2}}}}=\tilde C_0 a_{\tilde k}^2
\ee
and 
\be{tildeC0}
\tilde C_0=\frac{1}{2 k\pa{\sqrt{1+\frac{A_0}{2k^2}}}}\pa{\frac{2\nu-1}{-2\eta_0}}^{2\nu-1}\frac{\tilde k^{-2\nu+1}}{a_0^2}.
\ee
The perturbed solution in then given by
\be{persolPL}
p\rightarrow \frac{1}{2}\pa{\frac{q}{q-1}}^{\frac{2q}{q-1}} \pa{\eta_0}^{-\frac{2q}{q-1}}k^{-2\nu}\pa{\frac{a}{a_0}}^2\pa{1+\frac{A_0}{2k^2}}^{-\frac{q}{q-1}}
\ee
and the quantum gravitational corrections, which are enconded in the factor  $\pa{1+\frac{A_0}{2k^2}}^{-\frac{q}{q-1}}$, are negligible for large $k$. Let us note that this behaviour is simply dictated by the dependence of $A_0$ on $k$, that is , it is related to the dependence on $k$ of the unperturbed solution (in the de Sitter limit one correctly reproduces the $k^3$ dependence).

%%%%%%%%%%%%%%%%%%%%%%%%%%%%	
\subsection{Slow-Roll Inflation}
The de Sitter and the Power law evolutions are fairly good approximations to the inflationary dynamics. Furthermore these models permit an almost exact treatment of the primordial fluctuations and are thus of pedagogical interest. A wider class of more realistic inflationary models is that associated with the slow-roll dynamics. In such a case the evolution of cosmological perturbations occurs during a generic inflationary phase having a slowly varying Hubble parameter and a scalar field. The diverse inflationary models are then treated within the slow-roll (SR) approximation and the features of the spectra of perturbations, generated during inflation, are accurately estimated in such a framework, with an accuracy comparable with the magnitude of the so called SR parameters. It is then worth generalizing our procedure to such a case.\\
In the GR framework it is quite common to introduce the SR parameters
\be{SRlit}
\epsilon_{SR}\equiv -\frac{\dot H}{H^2}\;\;{\rm and}\;\; \eta_{SR}\equiv-\frac{\ddot \phi_0}{H\dot \phi_0}
\ee
and calculating the spectra just in terms of these two. The SR approximation consists of neglecting their derivatives (that is  treating them as constants) or, equivalently, to only keeping first order contributions in the SR variables . 

To first order in the SR approximations, the scale factor evolution satisfies the equation
\be{aSReq}
a H\simeq-\frac{1+\ep{1}}{\eta}
\ee
and its solution is then given by
\be{aSR}
a=a_0\pa{\frac{\eta_0}{\eta}}^{1+\ep{1}}.
\ee
In terms of the above quantities one finds
\be{omegas}
\omega^2=k^2-\frac{z''}{z}=k^{2}-\frac{2\pa{1+3\epsilon_{SR}-\frac{3}{2}\eta_{SR}}}{\eta^{2}}
\ee
for the scalar perturbation and 
\be{omegat}
\omega^2=k^2-\frac{a''}{a}=k^{2}-\frac{2\pa{1+\frac{3}{2}\epsilon_{SR}}}{\eta^{2}}
\ee
for the tensor perturbations. In contrast with the de Sitter and power-law cases, the equations for the scalar and the tensor perturbations are now different.  However, because of the forms of (\ref{omegas}) and (\ref{omegat}), it is possible to recover the equation/solution for the tensor perturbations starting from the equation/solution for the scalar perturbations and taking the limit $\eta_{SR}\rightarrow \ep{SR}$. We shall then focus on the scalar case and finally extract the tensor case results in the above limit. \\
We proceed in a fashion analogous to the power-law case. 
In the short wavelength regime, the quantum gravitational corrections evaluated perturbatively are absent at the leading and next to leading order. We thus neglect their contribution in such a limit. Conversely, in the long wavelength regime, the quantum gravitational correction should be taken into account and can be evaluated perturbatively. Finally the matching at the horizon crossing is performed. 

In the long wavelength regime, the quantum corrections may be rewritten as
\be{DPlong}
\Delta_p=\frac{ a^5 H^7}{k^6\Mt^2}\pa{7 \pa{\ep{SR}-\eta_{SR}}-4\frac{k^2}{a^2H^2}}\equiv \Delta_1+\Delta_2
\ee
where the first term
\be{D1}
\Delta_1\equiv \frac{7 a^5H^7}{k^6\Mt^2}\pa{\ep{SR}-\eta_{SR}},
\ee
is peculiar for the scalar sector in the SR case and the second term
\be{D2}
\Delta_2\equiv -\frac{4 a^3H^5}{k^4\Mt^2}
\ee
is common for de Sitter and Power Law cases. 
To the leading order, $p=C_0a^2\ep{SR}$ with $c_0=H_k^2/\pa{2k^3\ep{SR}}$, $p'/p=2aH$ and $p''/p=6a^2H^2$. The perturbed second order equation for $p$ is (\ref{peqper2}), where the integration on the r.h.s. is taken from $\eta_k=-1/k$ to $\eta$. \\
On integrating by parts one then finds
\be{contD1}
\frac{1}{p}\int_{-1/k}^\eta d\eta' p\Delta_1=\frac{A_0}{p}\int_{-1/k}^\eta d\eta'a^4H^6p'=\frac{A_0}{p}\pa{\frac{a^4H^6 p}{3}-\frac{k^3H_k^2}{6}}
\ee
with $A_0=\frac{7 \pa{\ep{SR}-\eta_{SR}}}{2k^6\Mt^2}$
and 
\be{contD2}
\frac{1}{p}\int_{-1/k}^\eta d\eta' p\Delta_2=-\frac{B_0}{p}\int_{-1/k}^\eta d\eta'a^2H^4p'=-\frac{B_0}{p}\pa{\frac{a^2H^4 p}{2}-\frac{kH_k^2}{4}}
\ee
with $B_0=\frac{4}{k^4\Mt^2}$.

The equation for $p$ then takes the following form:
\ba{peqperSR}
&&\paq{1+\frac{7}{18}\pa{\ep{SR}-\eta_{SR}}\frac{H^4}{H_k^2k^3\Mt^2}}p''-\frac{\pa{p'}^2}{2p}+2\pa{k^2-\frac{2H^4}{kH_k^2\Mt^2}-\frac{z''}{z}}p\nonumber\\
&&=\frac{1}{2p}\pa{1-\frac{2H_k^2}{k^3\Mt^2}}
\ea
and can be rewritten as
\be{peqperSR2}
\pa{1+\frac{\delta_k}{\Mt^2}}\tilde p''-\frac{\pa{\tilde p'}^2}{2\tilde p}+2\pa{\tilde k^2-\frac{z''}{z}}\tilde p=\frac{1}{2\tilde p}
\ee
with 
\be{defsolsr1}
\delta_k\equiv \frac{7}{18}\pa{\ep{SR}-\eta_{SR}}\frac{H_k^2}{k^3},
\ee
\be{defsolsr2}
\tilde k\equiv k\sqrt{1-\frac{2H_k^2}{k^3\Mt^2}}, \ee 
\be{defsolsr3}
\tilde p\equiv \pa{1-\frac{2H_k^2}{k^3\Mt^2}}^{-1/2} p
\ee  
where, on replacing $H\rightarrow H_k$, we neglected, to the leading order in SR, the time dependence of $H$.
The equation for $\tilde p$ is very similar to (\ref{peqper3}), except for the contribution proportional to $\delta_k/\Mt^2$. If $\delta_k/\Mt^2\ll 1$, which is consistent with our perturbative approach, one finds the following long wavelength solution for $\tilde p$ 
\be{solptildeSR}
\tilde p\simeq  \tilde C_0 z^{2\pa{1-\frac{\delta_k}{\Mt^2}}}
\ee
and consequently one has
\be{solpSR}
p=\tilde C_0\sqrt{1-\frac{H_k^2}{2k^3\M^2}} \,z^{2\pa{1-\frac{\delta_k}{\Mt^2}}}.
\ee
The integration constant $\tilde C_0$ is fixed by connecting the long wavelength solution to $p=1/2k$, when each mode $\tilde k$ crosses the horizon ($\tilde k=a_{\tilde k} H_{\tilde k}$). Finally one has
\be{psrfinal}
p=\frac{1}{2k}\paq{\frac{a^2H_k^2}{k^2\pa{1-\frac{H_k^2}{k^3\Mt^2}}}}^{1-\frac{7}{18}\pa{\ep{SR}-\eta_{SR}}\frac{H_k^2}{k^3\Mt^2}}
\ee
in the long wavelength regime and, given the smallness of the quantum gravitational corrections ($H_k/\M\ll 1$), one finally finds the expression
\be{psrfinalser}
p\simeq C_0 a^2\ep{SR}\paq{1+\frac{H^2}{k^3\Mt^2}\pa{1-\frac{7}{18}\pa{\ep{SR}-\eta_{SR}}\ln\frac{a^2H^2}{k^2}}}
\ee
valid for the scalar sector. In the tensor sector one easily obtains the corrections in the limit $\eta_{SR}\rightarrow \ep{SR}$. For such a case 
\be{psrfinalten}
p=\frac{a^2H^2}{2k^3\pa{1-\frac{H^2}{k^3\Mt^2}}}\ee
and
\be{psrfinalserten}
p\simeq C_0 a^2\ep{SR}\pa{1+\frac{H^2}{k^3\Mt^2}}.
\ee
%%%%%%%%%%%%%%%%%%%%%%%%%%%%%%
\section{Quantum gravitational corrections}
The effect of $\Delta_p$ on the evolution of the two-point function $p$ is that of adding to the standard, unperturbed, BD solution $p_{BD}$ a contribution of order $\Mt^{-2}$. When realistic inflationary models are considered, these modified spectra are derived from (\ref{psrfinalser}) and (\ref{psrfinalserten}) by replacing $k^3\rightarrow \pa{k/\bar k}^3$, where $\bar k$ is an unspecified reference wave number. The appearance of $\bar k=L^{-1}$ ,in the quantum corrections, can be traced back to the three volume integral in the original action for the homogeneous inflaton-gravity system plus perturbations (see the action (\ref{act})). Such a volume, on a spatially flat homogeneous space-time, is formally infinite and consequently the value of $\bar k$ remains undetermined. Naively one may argue that $\bar k$ is related to an infrared problem (divergence) and indeed, in the literature, its value is taken to be the infrared cut-off for the perturbations, namely the largest observable scale in the CMB. Alternatively one may consider it to be the scale at which new effects or physics  set in. We shall briefly return to this in the conclusions.\\
In the previous section we calculated the form of the quantum gravitational modifications to the primordial scalar spectrum, in the case of SR inflation
\be{QGCmod}
Q_k=1+\frac{H^2\bar k^3}{\Mt^2 k^3}\pa{1-\frac{7}{18}\pa{\ep{SR}-\eta_{SR}}\ln\frac{a^2H^2}{k^2}}.
\ee
In such an expression, the wavenumber $k$ necessarily refers to the scales, around the pivot scale $k_*$, which are probed by the CMB and exited from the horizon $N_*\sim 60$ e-folds before inflation ends. Its contribution to (\ref{QGCmod}) is
\be{keta}
\pa{\frac{k}{a H}}^{-2\pa{\ep{SR}-\eta_{SR}}}\simeq\pa{\frac{k_*}{a_*{\rm e}^{N*}H_{k_*}}}^{-2\pa{\ep{SR}-\eta_{SR}}}\simeq{\rm e}^{2N_*\pa{\ep{SR}-\eta_{SR}}}
\ee 
and may well lead to a contribution of $\mathcal{O}\pa{1}$ for reasonable values of the SR parameters of the order of 1 per cent. Let us note that the first equality, in (\ref{keta}), is strictly valid for the modes very close to the pivot scale $k\sim k_*=a_* H_{k_*}$. Away from the pivot scale, small deviations proportional to the SR parameters, $-2(\ep{SR}-\eta_{SR})\ln \pa{\frac{k}{k_*}}$, are neglected. 
Depending on the SR parameters and on $N_*$, the quantum corrections $Q_k$ may lead to a power loss or a power increase for large scales which can be generically parametrized in the following form:
\be{pparam}
p^{(L)}\simeq p_{0}^{(L)}\paq{1\pm q\pa{\frac{k_*}{k}}^3}
\ee
where $p_0^{(L)}$ is $p$ without quantum corrections and evaluated in the long wavelength regime. The quantity inside the square brackets is $Q_k$.
An analogous parametrization holds for the tensor sector with a different $q$.\\
%%%%%%%%%%%%%%%%%%%%%%%%%%
\subsection{Extrapolation beyond NLO}
The parametrization of the primordial spectra by (\ref{pparam}) is still not suitable for comparison with observations. In the $k\ll k_*$ limit the quantum gravitational corrections are either negative or very large (infinite in the $k\rightarrow 0$ limit). Such an apparently pathological behavior is simply a consequence of the perturbative technique employed to evaluate the corrections. 
%In principle, beyond the next to leading order, the perturbative series is convergent and the infrared limit of the power spectrum is regular. 
One may hope that resummation to all orders leads to a finite result.
In any case we are not allowed to extend the validity of the perturbative corrections up to $\mathcal{O}\pa{1}$.\\
Thus, instead of introducing a sharp cut-off on the NLO expressions for the modified spectra by multiplying $q$ by an ad hoc step function which keeps the correction small but leads to a discontinuous spectrum, we interpolate our expression through a well defined function, with a finite and reasonable behavior in the $k\rightarrow 0$ limit. Such a function, which must reproduce (\ref{pparam}) when $q\pa{k_*/k}^3\ll 1$, may be regarded as a resummation of the perturbative series.\\
In order to restrict the number of parameters which will be fitted by the comparison with the data and still allow for different limits when $k\rightarrow 0$ ,we consider the following parametrization:
\be{extparam}
p^{(L)}\simeq p_{0}^{(L)}\frac{1+\tilde q_1\pa{\frac{k_*}{k}}^3}{1+\tilde q_2\pa{\frac{k_*}{k}}^3}\sim p_{0}^{(L)}\paq{1+\pa{\tilde q_1-\tilde q_2}\pa{\frac{k_*}{k}}^3}.
\ee
where one more parameter w.r.t. (\ref{pparam}) has been added, in order to obtain a regular expression for $k$ small. 
Let us note that the above modifications are substantially different from considering a running spectral index $\alpha_s$, such as
\be{running}
p^{(L)}\simeq p_{0}^{(L)}\pa{\frac{k_*}{k}}^{-\frac{\alpha_s}{2}\ln\pa{\frac{k}{k_*}}}.
\ee
Indeed for the latter case, the standard power law dependence is affected at both large and small scales and, in particular, a negative running would lead to a zero amplitude in the $k\rightarrow 0$ limit and a smaller amplitude w.r.t. simple power law when $k\gg k_*$. On the other hand the modified spectrum (\ref{extparam}) reduces to the power law case when $k\gg k_*$ and may lead to a non zero amplitude when $k\rightarrow 0$, depending on the choice of the parameters $\tilde q_{1,2}$. 
%%%%%%%%%%%%%%%%%%%%%%%
\begin{table}[t]
\caption{Range of parameters varied }
\vspace{0.1 cm}
\centering
\begin{tabular}{c | c | c | c | c |  c | c }
\hline\hline
$\tau$& $\ln\pa{10^{10}A_s}$ & $n_s$ & $r$ & $\alpha_s$ & $q_1$ & $q_2$  \\ 
[1ex]
\hline
 $[0.01,0.8]$	&$[2.7,4.0]$		& $[0.9,1.1]$ 		& [0,0.8] 		& [-0.1,0.1] 		& [0,21] 		& [0,0.5]			\\ 
[1ex]
\hline
\end{tabular}
\label{tab1}
\end{table}

\section{Data Analysis}
In this section we report the comparison between the theoretical predictions given by (\ref{extparam}) and Planck 2015 \cite{planck} dataset. The analysis is performed using the Markov Chain Monte Carlo (MCMC) code {\it COSMOMC} \cite{cosmomc}, which has been properly modified to take into account the estimated quantum gravitational effects.\\
Let us note that BD vacuum in the tensor sector gives a power increase for large scales in the tensor spectrum. Such an increase would be counterbalanced by a loss of power in the scalar sector, as far as temperature correlations are concerned. One may parametrize such a power increase in a suitable way, just as we did for the scalar sector ,in order to eliminate the divergence for small $k$ and fit the corresponding parameter with the data at our disposal. Since our main source of data comes from temperature correlations, which do not discriminate between scalar and tensor fluctuations, we neglect a priori quantum gravitational corrections in the tensor spectrum. Such a choice is a simplifying assumption done in order not to have to disentangle possible degenerate parameters. Let us note, however, that such a choice can be realized physically either by an appropriate vacuum choice, differing from a pure BD, or by a very long cutoff scale associated with tensor dynamics. Thus we limit our analysis to a subset of the more general case, for which the quantum gravitational corrections affect the tensor sector in a non negligible way, thus minimizing the power loss in the scalar sector. The tensor spectrum is then given by the unperturbed power law expression
\be{tensor}
p_t=A_t\pa{\frac{k}{k_*}}^{n_t}.
\ee
and we assume that the LO spectra are generated by the conventional SR mechanism and single field inflation. The consistency condition, relating scalar and tensor spectral indices and the tensor to scalar ration, is valid when quantum gravitational corrections are neglected. Indeed throughout the analysis we assume that the consistency relation (already implemented in  {\it COSMOMC}) between the spectral indices and the tensor to scalar ratio
\be{consistency}
n_t=-\frac{r}{8}\pa{2-n_s-\frac{r}{8}}
\ee
holds to the second order in the SR approximation and the amplitude of the spectrum of tensor perturbations is given by $A_t=r A_s$, to the leading order in $\M^{-1}$, i.e. on neglecting the quantum gravitational corrections. 
We then consider a primordial scalar spectrum $p^{(L)}$ parametrized by 
\be{scparam}
p_s\simeq p_{0}^{(L)}\frac{1+\pa{1-2q_2}\pa{\frac{k_*}{{\rm e}^{q_1}k}}^3}{1+\pa{\frac{ k_*}{{\rm e}^{q_1}k}}^3}
\ee
where $1-2q_2$ simply fixes the limit of $p_s$ when $k\rightarrow 0$. The parameter $q_1$ is related to the scale $\bar k$, i.e. that at which the quantum gravitational modifications of the spectrum become important. In the limit $q_1\rightarrow \infty$ ($\bar k\rightarrow 0$), the quantum gravitational corrections are suppressed and for $q_1=0$ one has $\bar k=k_*$. Let us note that $q_2=0$ ,or $q_1\rightarrow \infty$, correspond to the standard power-law case with no loss of power ($p_s=p_{0}^{(L)}$) and $q_2=0.5$ corresponds to zero power at $k=0$. The expression (\ref{scparam}) is a parametrization equivalent to (\ref{extparam}), with $\tilde q_1=\exp{\pa{-3q_1}}\pa{1-2q_2}$ and $\tilde q_2=\exp{\pa{-3q_1}}$, which we have found to be more convenient to be used in {\it COSMOMC}.\\
\begin{table}[t]
\caption{List of Models}
\vspace{0.1 cm}
\centering
\begin{tabular}{c | l | |l|l}
\hline\hline
Model \# & Primordial spectra & Datasets & Parameters  \\ 
[1ex]
\hline
1	 			& Power law 		 	& PL &$A_s$, $n_s$, $r$		\\
2		 		& and tensors	 	& PL+BK 	&	\\
\hline
3			 	& Running spectral index 	 	& PL	&$A_s$, $n_s$, $r$, $\alpha_s$	\\
4			 	& and tensors	 	& PL+BK			&	\\
\hline
5			 	& Quantum gravitational 	 	&PL 		&$A_s$, $n_s$, $r$, $q_1$, $q_2$	\\
6				&  corrections and tensors		&PL+BK	& \\ 
[1ex]
\hline
\end{tabular}
\label{tab2}
\end{table}
Our analysis is based on the Planck datasets released in 2015 and includes the Planck TT data with polarization at low $l$ (PL), and the data of the BICEP2/{\it Keck Array}-Planck joint analysis (BK) \cite{bicep}. In particular we use {\it  plik\_dx11dr2\_HM\_v18\_TT}, {\it lowTEB} and {\it BKPlanck} publicly available Planck likelihoods. We find the best fit for our model with and without BK data and compare it with standard power law predictions, and with those assuming a non negligible running of the spectral index (\ref{running}).\\
For simplicity we obtained the best-fits for the parameters of the primordial spectra shown in Table \ref{tab1} and the parameters are taken to vary with uniform priors in the intervals indicated in the same table. The priors for $\tau$, $A_s$, $n_s$, $r$ and $\alpha_s$ are those used by the  Planck 2015 analysis. The remaining cosmological parameters are fixed to the Planck best-fit and in particular we chose
\be{planckfix}
100\theta_{MC}=1.040,\; \Omega_bh^2=0.0222,\; \Omega_c h^2=0.119
\ee

Let us note that the pivot scale $k_*$ is $0.05\;{\rm Mpc}^{-1}$ and is the same for both the scalar and the tensor sector.  The additional parameters $q_1$ and $q_2$ are chosen to vary in the largest possible interval leading to a power loss for large scales (compared with the pivot scale), with the parametrization chosen. At present our theoretical predictions are not able to constraint the value of such parameters, or estimate possible allowed intervals where to let them vary (see \cite{hert} for an attempt to estimate priors from quantum gravity), thus the choice of broad enough priors seems reasonable.\\	
In particular the prior for $q_2$ is chosen to let it vary between $q_2=0$, where the quantum gravitational corrections cancel out independently of $q_1$, and $q_2=1/2$. The values for $q_2$ with $q_2>1/2$, lead to an increase of power, those with $q_2<0$, lead to a physically unacceptable negative spectrum and are thus excluded from the analysis.\\
The choice of the prior for $q_1$ is rather delicate with the parametrization chosen (\ref{scparam}). On expanding (\ref{scparam}) to the first order in the quantum gravitational corrections and comparing the result with the theoretical predictions (\ref{QGCmod}), one finds, after some algebra, the following relation among the parameters of our model
\be{q1prior}
\exp{\pa{3q_1}}= \frac{24\,q_2 }{\pi^2\,r\cdot A_s \cdot Q(n_s,r,N_*)}\pa{\frac{k_*}{\bar k}}^3 
\ee
with 
\be{defQ}
Q(n_s,r,N_*)\equiv \frac{7}{18}\pa{1-n_s-\frac{r}{8}}N_*-1
\ee
where we have used the following standard SR relations for single field inflation:
\be{scaleinf}
\frac{H_*^2}{\M^2}\simeq \frac{\pi^2}{2}A_s\cdot r
\ee
and
\be{srdata}
r=16\,\ep{SR},\; n_s=1+2\,\eta_{SR}-4\,\ep{SR}.
\ee

\begin{figure}[t!]
\centering
\epsfig{file=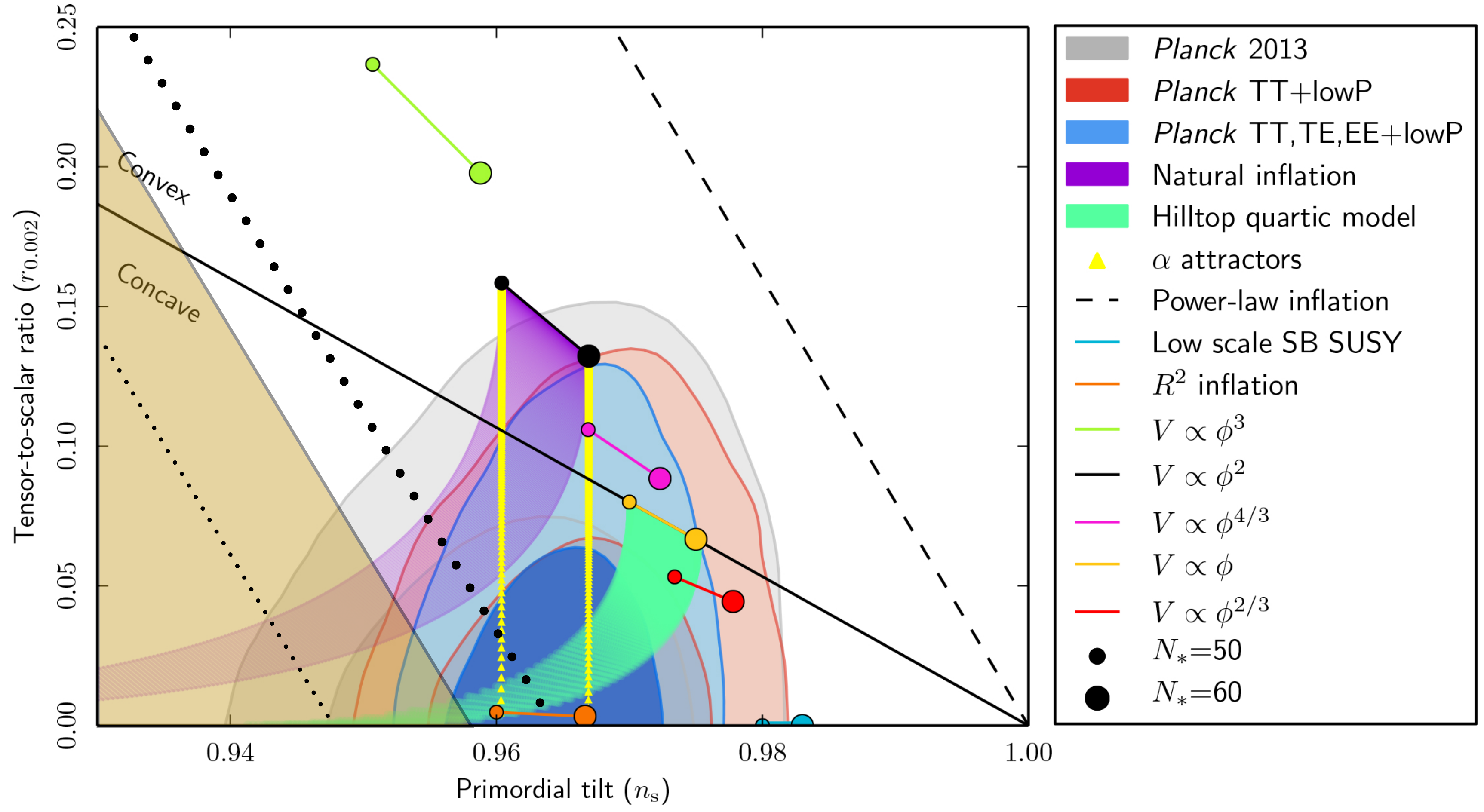, width=13 cm}
\caption{\it The figure plots the region (yellow area) compatible with a loss of power in the $(n_s,r)$ plane for $N_*=60$. The dotted lines are the contours of the $N_*=50$ (small dots) and $N_*=70$ (large dots) areas. These contours are superimposed on the  Planck 2015 analysis of various inflationary models. Hilltop quartic models and Natural inflation models lead to a loss in power, conversely chaotic inflation is not compatible with such a loss. 
\label{plcomp}}
\end{figure}
The prior for $q_1$ then depends on some other priors of observables quantities (such as $r$ and $n_s$) and on a  few, related, physical assumptions. \\
Let us first note that, with the priors considered for the quantities on the r.h.s. of (\ref{q1prior}), such expression may vary from $-\infty$ to $+\infty$. The case of power loss, which we are investigating, is only reproduced by positive values of $Q$ and ,correspondingly, the r.h.s. of (\ref{q1prior}) then varies in the interval $\paq{0,+\infty}$ (let us note that the relation (\ref{q1prior}) is otherwise undefined).  Such a positivity requirement can be fulfilled only by particular inflationary models, as figure (\ref{plcomp}) shows, with larger values of $N_*$ generically favoured compared to smaller ones.
For example consider the case of chaotic inflation, driven by a power-law potential $V\propto \phi^n$. For such a case 
\be{rnspower}
\pa{n_s,r}=\pa{1-\frac{2\pa{n+2}}{4N_*+n},\frac{16n}{4N_*+n}}\stackrel{N_*\gg n}{\longrightarrow}\pa{1-\frac{n+2}{2N_*},\frac{4n}{N_*}},
\ee
\be{chaotic}
Q\sim -\frac{11}{18}
\ee
and (\ref{q1prior}) is undefined. \\
Conversely for the Hilltop inflationary models, one has 
\be{rnshill}
\pa{n_s,r}=\pa{1-\frac{2\pa{n-1}}{N_*\pa{n-2}},\frac{1}{\pa{N_*}^{-2\frac{n-1}{n-2}}}},
\ee
where $n$ is defined by the shape of the potential
\be{hillpot}
V_{\rm hilltop}=V_0\paq{1-\pa{\frac{\phi}{\mu}}^n}
\ee
and
\be{hilltop}
Q\sim -\frac{11-2n}{18-9n}
\ee
leading to a loss in power for $2<n<11/2$.\\
Let us note that, with the form obtained for the quantum gravitational corrections, our model leads to severe constraints on the shape of the inflationary potential. As shown in the figure (\ref{plcomp}), only a small subset of the inflationary models, satisfying the observed values of  $n_s$ and $r$, lead to a loss of power for large scales. The remaining models would give a power increase, which may be a distinguishing feature, unless $\bar k$ is too small to be observed in the CMB. More generally, on referring to the classification in \cite{classinf},  power loss is associated only with a sub set of class I models, with $n_s\simeq 1+2\ln{b}/N_*$, $r\propto 1/N_*^{-2\ln{b}}$ and $0<b<0.277$ for $50<N_*<70$.

For the models which lead to  a loss in power we assume 
\be{pr1}
-\frac{1}{Q(n_s,r,N_*)}\sim \mathcal{O}(10)
\ee
which is the order of magnitude of (\ref{hilltop}), far from the boundaries $2$, $11/2$.

The ratio $k_*/\bar k$, where $\bar k^{-1}$ is the scale at which the power loss begins to be observable, is taken in the interval $\paq{10^{-1},10^3}$, where $10^{-1}$ is the order of magnitude of the shortest scales probed by Planck and $10^3$ corresponds to largest scale one can observe (in units of the pivot scale). The tensor to scalar ratio $r$, appearing at the denominator, in principle can be $0$, as we vary it in the interval $\paq{0,0.8}$. However, on attempting to provide a finite prior for $q_1$ (and only in this context), we observe that most single field inflationary models generate a non zero tensor to scalar ratio and, therefore,  we shall  assume it varies in the interval $\paq{10^{-4},0.8}$ where $10^{-4}\sim \mathcal{O}\pa{1/N_*^2}$. Similarly $q_2$ is taken in the interval $[10^{-2},1/2]$, where $10^{-2}$ corresponds to a $2\%$ power loss, as smaller values of $q_2$ would be indistinguishable from zero and lead to an infinite prior for $q_1$. Finally  the amplitude $A_s$ varies in the interval $\paq{1.5\cdot 10^{-9}, 5.5\cdot 10^{-9}}$. Given all such assumptions ,the prior for $q_1$ can be estimated to be $\paq{0,21}$.

The different combinations of primordial spectra and datasets considered, are listed in Table \ref{tab2} with an index specifying the model number. The best fits found, for the parameters we varied ,are presented in Table \ref{tab3} and the corresponding effective $\chi^2$, defined as $-2\ln\mathcal{L}$ where $\mathcal{L}$ is the likelihood, are listed in Tables \ref{tab4} and \ref{tab4b}. The differences between the total $\chi^2$ for the different cases are reported, using our model as reference. In particular the cases 1 and 3 are compared with 5 and the cases 2 and 4 are compared with 6.
\begin{table}[t]
\caption{Monte Carlo Best-fits}
\vspace{0.1 cm}
\centering
\begin{tabular}{c | c c c c c | c c }
\hline\hline
\# & $\tau$& $\ln\pa{10^{10}A_s}$ & $n_s$ & $r$ & $\alpha_s$ & $q_1$ & $q_2$  \\ 
[1ex]
\hline
1		&$7.7\cdot 10^{-2}$	& 3.09 	& 0.965 		& $1.05 \cdot10^{-2}$ 		& - 					& -					& -						\\
2	 	&$8.3\cdot 10^{-2}$ 	& 3.10 	& 0.967 		& $1.65\cdot 10^{-2}$ 		& - 					& -					& -						\\
3		&$7.8\cdot 10^{-2}$	& 3.09 	& 0.964 		& $1.85 \cdot 10^{-2}$ 		& $-1.02\cdot 10^{-2}$ 	& -					& -					   	\\
4		&$8.9\cdot 10^{-2}$	& 3.11 	& 0.967		& $3.13 \cdot 10^{-2}$ 		& $-6.65\cdot 10^{-3}$	& -					& -						\\
5		&$8.0\cdot 10^{-2}$	& 3.09 	& 0.965 		& $1.63\cdot 10^{-2}$ 		& - 					& $3.48$	& $1.3\cdot 10^{-1}$					\\
6		&$8.9\cdot 10^{-2}$	& 3.12 	& 0.966 		& $4.7\cdot 10^{-2}$ 			& -	  				& $2.64$	& $5.6 \cdot 10^{-2}$				 	\\ 
[1ex]
\hline
\end{tabular}
\label{tab3}
\end{table}
%%%%%%
\begin{table}[ht]
\caption{Monte Carlo Comparison (PL)\label{tab4}}
\vspace{0.1 cm}
\centering
\begin{tabular}{c | c | c }
\hline\hline
\# & $\chi^2_{Tot}$ & $\Delta \chi^2\equiv \chi^2_\#-\chi^2_7$  \\
[1ex]
\hline
1		 				&  11265.3		& 3.3     	\\
3		 	 			&  11265.1   		& 3.1	\\
\hline
5			 			&  11262.0	 	& 0		\\
\hline
\end{tabular}
%\caption{\it $\chi^2$ values for the best fits with the models listed in the first column. In the last column the $\chi^2$ differences between each model`s best fit and the reference model best fit ($\# 5$) is given.}
\end{table}
\begin{table}[ht]
\caption{Monte Carlo Comparison (PL+BK)\label{tab4b}}
\vspace{0.1 cm}
\centering
\begin{tabular}{c | c | c }
\hline\hline
\# & $\chi^2_{Tot}$ & $\Delta \chi^2\equiv \chi^2_\#-\chi^2_8$ \\
[1ex]
\hline
2	 				&  11307.4		& 4.1	\\
4		 			&  11307.3 		& 4.0	\\
\hline
6			 		&  11303.3 		& 0		\\ 
\hline
\end{tabular}
%\caption{\it $\chi^2$ values of the best fits for models listed in the first column. In the last column the $\chi^2$ differences between each model best fit and the reference model best fit ($\# 6$) is reported. \label{tab4b}}
\end{table}

%%%%%%%%%%%%%%%%%%%%%
\subsection{Results}
The MCMC results (see Tables \ref{tab4} and \ref{tab4b}) show that the quantum gravitational modification of the standard power law form for the primordial scalar spectrum, improves the fit to the data. Such improvements are much more significant w.r.t the standard modifications of the primordial spectra obtained on considering a running spectral index. Let us note that the 2015 Planck data give constraints on the running, which are quite different from those coming from the  2013 data. In particular the fit to the  2015 data does not improve much if one considers a running spectral index in the scalar sector.\\
\begin{figure}[t!]
\centering
\epsfig{file=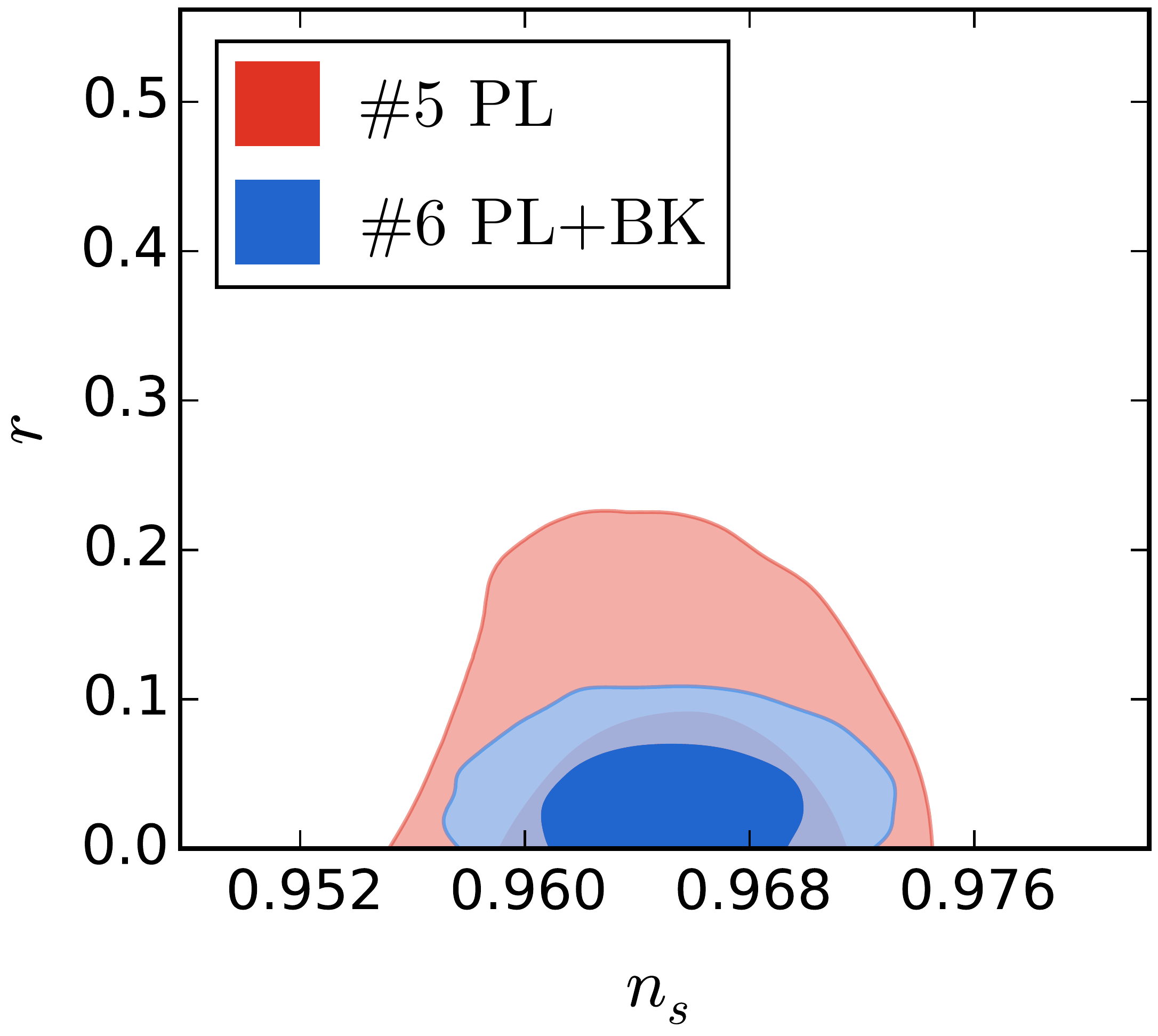, width=8 cm}
\caption{\it The figure shows the $68\%$ and $95\%$ confidence level constraints on $r$ and $n_s$.}
\label{fig1b}
\end{figure}
The comparison of the marginalized 1-D likelihoods for the parameters $q_1$ and $q_2$ in  Fig. (\ref{fig3}) show that the two datasets lead to close predictions. In particular their marginalized maxima are
\be{bestfitq}
q_1\simeq 3.4\;,\;\;2q_2\simeq 0.23
\ee
when Planck data alone are considered and
\be{bestfitq2}
q_1\simeq 3.8\;,\;\;2q_2\simeq 0.20
\ee
when BK data are added to the analysis. 
Correspondingly $n_s$ and $A_s$ also take very similar values for the best fit. 
%BICEP2 data restrict the allowed interval for $r$ leading to results similar to the running case.

The value of $q_2$ indicates a $\sim 20-25\%$ loss in power when $k$ approaches zero.  Let us note that the tensor to scalar ratio $r$ is weakly constrained.\\
From tables (\ref{tab4}-\ref{tab4b}) we observe that  cases 3 and 4, with a running spectral index, are disfavoured w.r.t cases 1 and 2 respectively, since they almost the same, effective, $\chi^2$ , but with one more independent d.o.f. to fit the data. Conversely the cases 5 and 6 ($\Delta\chi^2> 2$) are favoured w.r.t. the cases 1-4, as an improvement greater than 2 for the effective $\chi^2$ is obtained, through the addition of 2 independent parameters. \\
In figure \ref{fig3} we finally plot the marginalized likelihoods for $r$, $q_1$ and $q_2$. The corresponding marginalized $68\%$, $95\%$ and $99\%$ confidence intervals are listed in Tables \ref{tab7}-\ref{tab8}.
The marginalized likelihoods for $q_1$ and $q_2$ show a $1\sigma$ deviation from standard power law for both cases 5 and 6. Let us note that, on comparing the results with  those obtained from the Planck 2013 data, the constraints on $q_1$ and $q_2$ are now weaker \cite{K}.\\

Finally let us discuss the constraint on $\bar k$.
On assuming, for example, Hilltop inflation (\ref{rnshill}), one can invert the relation (\ref{q1prior})  obtaining
\be{fitphys}
\frac{\bar k}{k_*}\simeq \exp{\pa{-q_1}} \pa{\frac{24\,q_2\,N_*^{2\frac{n-1}{n-2}}}{\pi^2\, A_s}\,\frac{9n-18}{11-2n}}^{1/3}.
\ee

\begin{figure}[t!]
\centering
\epsfig{file=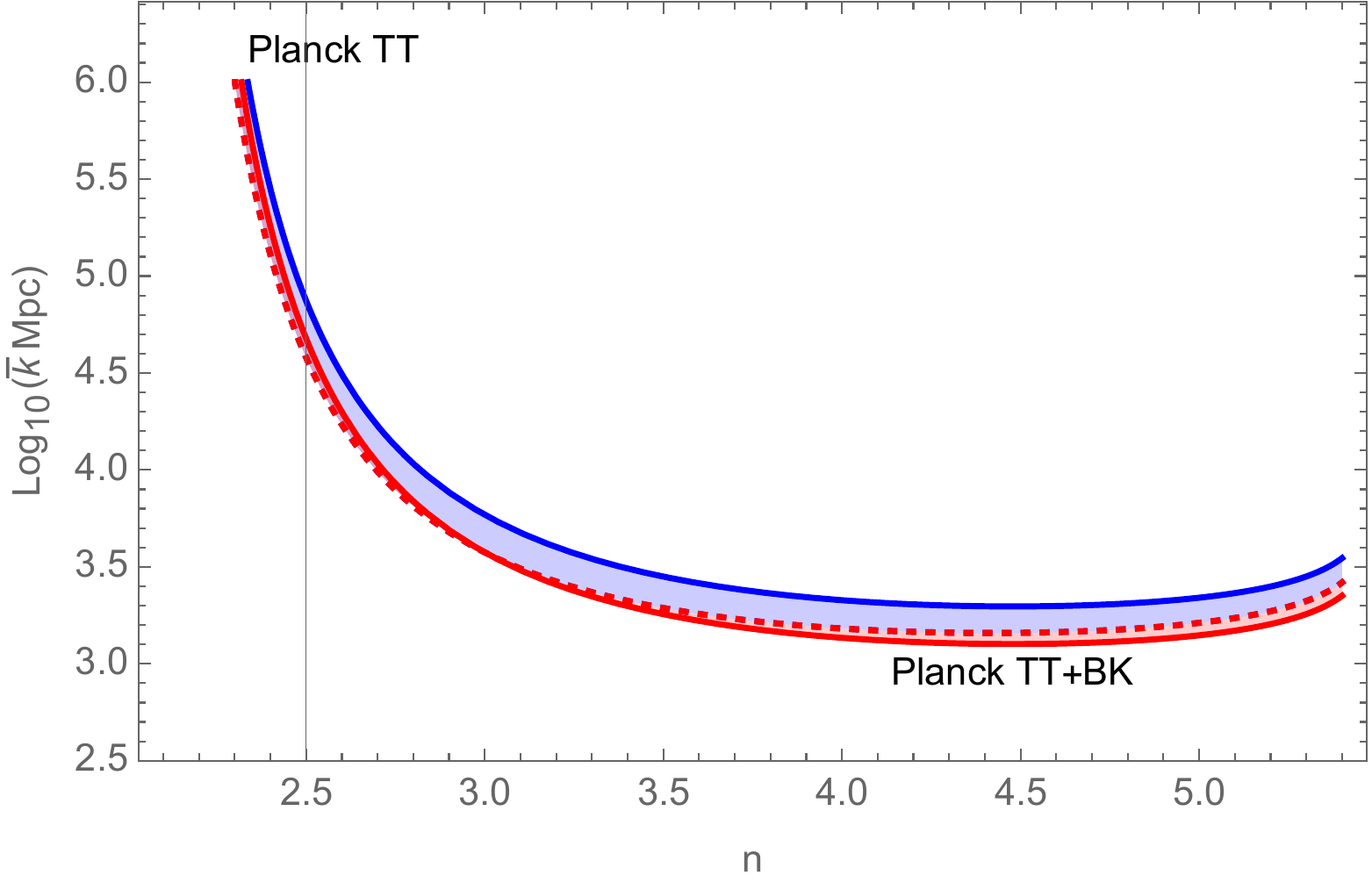, width=7 cm}
\caption{\it Constraints on $\bar k$ for hilltop inflation as a function of $n$ without (blue region) and with (red region) BK data. The region spans different $N_*\in \paq{50,70}$.}
\label{barkhill}
\end{figure}

\begin{figure}[t!]
\centering
\epsfig{file=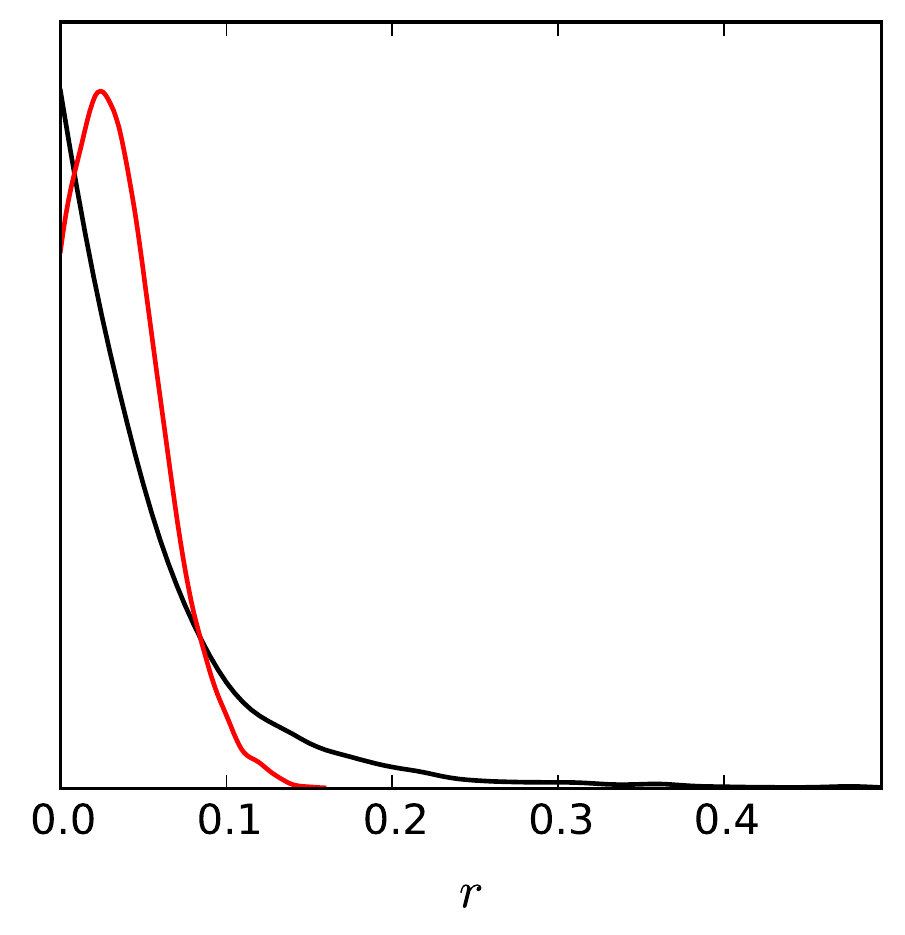, width=4.4 cm}
\epsfig{file=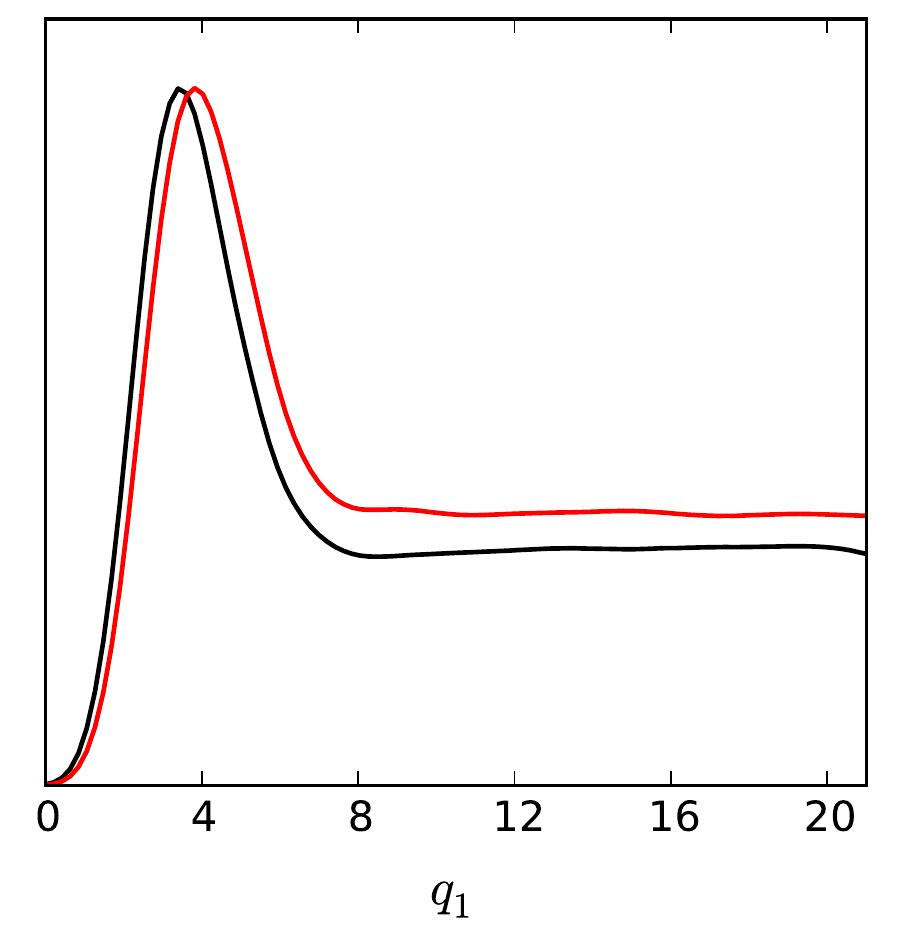, width=4.4 cm}
\epsfig{file=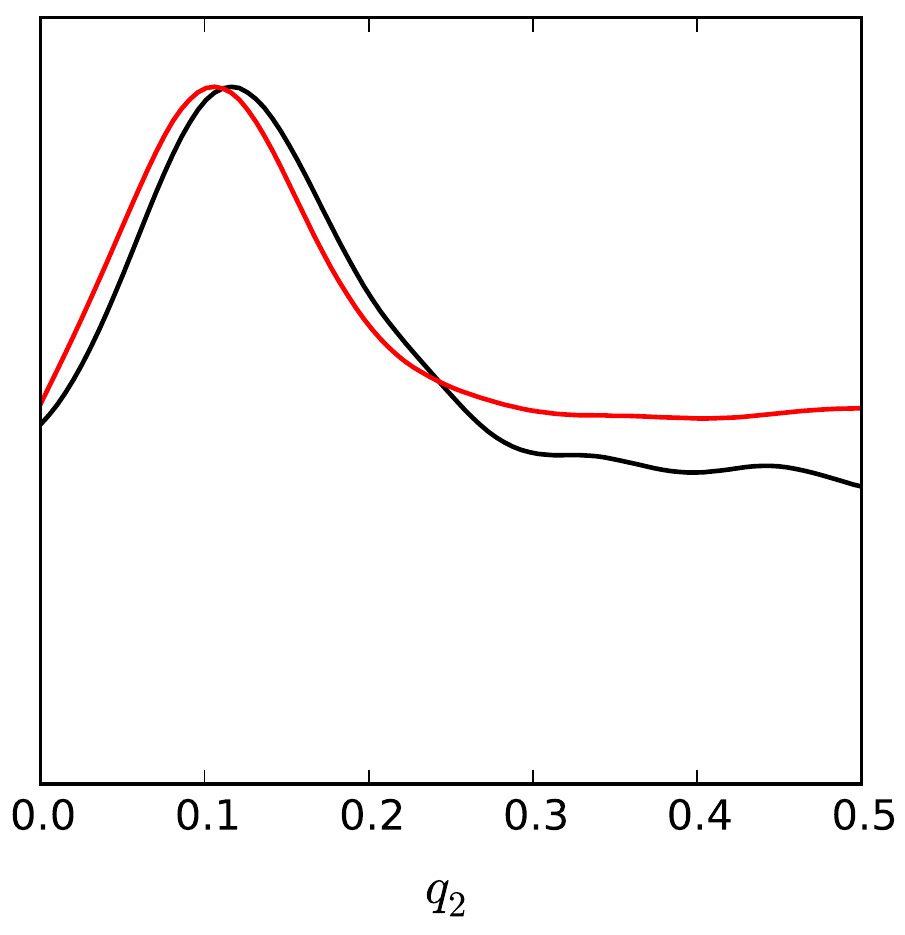, width=4.4 cm}
\caption{\it Marginalized 1-D likelihoods for $r$, $q_1$ and $q_2$ without (black line) and with (red line) BK data.}
\label{fig3}
\end{figure}

Given that the amplitude $A_s$ is quite constrained by observations and, on using $50<N_*<70$, $n=4$, we obtain the corresponding values for $\bar k$, which are very large compared to the wave number associated with the largest observable scale in the CMB namely $k_{\rm min}\simeq 1.4\cdot  10^{-4}\;{\rm Mpc}^{-1}$. These values are illustrated in  figure (\ref{barkhill}) for the cases $\#$ 5 and 6 as functions of $n$ (defined by (\ref{hillpot})).
Let us note that the existence of such a (relatively) small fundamental length may have relevant consequences on astrophysical observation. Indeed it is associated with distances which are comparable with the diameter of a large galaxy or a galaxy cluster. We further observe that a 3 order of magnitude variation of the value of $\bar k$ can be obtained on ``re-tuning'' the parameters used for its estimate. Let us note that the estimate for $\bar k$, although illustrated for a specific inflationary model, is quite general and can also  be found for other diverse power loss compatible models. \\

%In figure \ref{fig2} we compare our best fits for the cases $2$, $4$, $6$, $8$ with the observed Planck+WMAP $C_l^{TT}$ (for $l<50$ where differences emerge w.r.t. the standard power law spectra) and BK $C_l^{BB}$. 
%Let us note that, as far as the best fits are concerned, cases $7$ and $8$ give very similar predictions. 

\begin{table}[h!]
\caption{Marginalized confidence intervals - Case 5}
\vspace{0.1 cm}
\centering
\begin{tabular}{c | c c c}
  & $68\%$ & $95\%$& $99\%$  \\
[0.1ex]
\hline
$r$			& $[0.0,6.2\cdot 10^{-2}]$ 	& $[0.0,1.8\cdot 10^{-1}]$ 			&  $[0.0,3.2\cdot 10^{-1}]$ 	\\
$q_1$		& $[1.7,2.0 \cdot 10^1]$	& $[2.5,2.1\cdot 10^1]$ 	& $[2.1.2.1\cdot 10^1]$	\\ 
$q_2$		& $[0.0,2.8\cdot 10^{-1}]$	& $[0.0,5.0\cdot 10^{-1}]$ 	& $[0.0,5.0\cdot 10^{-1}]$	\\
[1ex]
\hline
\end{tabular}
\label{tab7}
\end{table}
\begin{table}[h!]
\caption{Marginalized confidence intervals - Case 6}
\vspace{0.1 cm}
\centering
\begin{tabular}{c | c c c}
  & $68\%$ & $95\%$& $99\%$  \\
[0.1ex]
\hline
$r$			& $[0.0,4.8\cdot 10^{-2}]$ 	& $[0.0,8.4\cdot 10^{-2}]$ 			&  $[0.0,1.2\cdot 10^{-1}]$ 	\\
$q_1$		& $[2.1,1.6\cdot10^1]$	& $[2.7,2.1\cdot 10^1]$ 	& $[2.2,2.1\cdot 10^1]$	\\ 
$q_2$		& $[0.0,5.0\cdot 10^{-1}]$	& $[0.0,5.0\cdot 10^{-1}]$ 	& $[0.0,5.0\cdot 10^{-1}]$	\\
[1ex]
\hline
\end{tabular}
\label{tab8}
\end{table}

%%%%%%%%%%%%%%%%%%%%%%%%%
\section{Conclusions}
As we mentioned in the introduction the matter-gravity system is amenable to a Born-Oppenheimer treatment, wherein gravitation is associated with the heavy (slow) degrees of freedom and matter with the light (fast) degrees of freedom. Once the system is canonically quantised and the associated wave function suitably decomposed, one obtains that, on neglecting terms due to fluctuations (non-adiabatic effects), in the semiclassical limit gravitation is driven by the mean matter Hamiltonian  and matter follows gravitation adiabatically, while evolving according to the usual Schwinger-Tomonaga (or Schr\"odinger)  equation. Our scope in this paper has been to study perturbatively the effect of the non-adiabatic contributions,  for different inflationary backgrounds. In particular we wished to see such effects on the observable features of the scalar/tensor fluctuations generated during inflation. In order to do this we obtained a master equation for the two-point function for such fluctuations, which includes the lowest order  quantum  gravitational corrections. These corrections manifest themselves on the largest scales, since the associated perturbations are more effected by quantum gravitational effects, as they exit the horizon at the early stages of inflation and are exposed to high energy and curvature effects for a longer period of time. Interestingly the very short wavelength part of the spectrum remains unaffected and one may consistently assume the BD vacuum as an initial condition for the evolution of the quantum fluctuations. Computationally this feature is relevant as it allows one to find the long wavelength part of the spectrum of the fluctuation through a matching procedure (similar to the standard case without quantum gravitational corrections). \\
In particular one finds, for a de Sitter evolution, a power enhancement w.r.t. the standard results for the spectrum at large scales, with corrections behaving as $k^{-3}$. Such a $k^{-3}$ was also found with similar approaches \cite{quantumloss} and may appear to be a peculiarity of such quantum gravity models. However the case of power law inflation is different: while power enhancement is also true for power-law inflation, of interest for this case is that one finds that the $k$ dependence of the quantum gravitational corrections differs from $k^{-3}$ and is, perhaps not surprisingly, directly related to the $k$ dependence of the unperturbed spectra.\\
Finally it is the slow roll case that is more realistic and of greatest interest. The quantum gravitational corrections for the SR case have peculiar features and are very different from the de Sitter case. In particular, for the case of the scalar fluctuations, their form is not simply a deformation of the de Sitter result proportional to the SR parameters. New contributions arise due to SR and their effect is comparable with the de Sitter-like contributions for very large wavelengths. The new contributions are proportional to $\ep{SR}-\eta_{SR}$ and are zero for the de Sitter and power-law cases. They can lead to a power-loss term for low $k$ in the spectrum of the scalar curvature perturbations at the end of inflation, providing the difference $\ep{SR}-\eta_{SR}>0$. The evolution of the primordial gravitational waves has also been addressed. The quantum gravitational corrections also affect the dynamics of tensor perturbations and determine a deviation from the standard results in the low multipole region, which always leads to a power enhancement.
In performing the analysis, for simplicity,  we restricted  ourselves to the particular case of negligible quantum gravitational contributions to the spectrum of primordial gravitational waves. Further, since our corrections are perturbative,in order to keep  them so  for  all values of k, we have suitably extrapolated our predictions  for the scalar sector beyond the leading order, describing this  in terms of two parameters, and examined them down to $k\rightarrow 0$. Other parametrizations have also been considered, however the one we presented is the simplest and leads to the best results.\\ 
%Further we assumed that the LO spectra are generated by the conventional SR mechanism and single field inflation. As a consequence the consistency condition relating scalar and tensor spectral indices and the tensor to scalar ration are valid to such a LO. Different potentials lead to differing expressions for these quantities. 
It is found  that, given the form obtained for the quantum gravitational corrections, our model imposes severe constraints on the shape of the inflationary potential ,as a loss in power at large scales is compatible with observations, whereas a power enhancement must be zero or extremely small to fit the data. Only a small subset of the inflationary models, satisfying the observed values for $n_s$ and $r$, lead to a loss of power at large scales. The remaining models give a power increase which may be a distinguishable feature, unless $\bar k$ is too small to be observed in the CMB.\\
Finally the analysis performed was  based on Planck datasets released in 2015 , include the Planck TT data with polarization at low $l$ (PL) and the data of the BICEP2/{\it Keck Array}-Planck joint analysis (BK) \cite{bicep}. In our preceding paper \cite{K} our model predictions were tested through Planck 2013 and BICEP2 earlier data and the results were different.
The MCMC results (see Tables \ref{tab4} and \ref{tab4b}) show that the quantum gravitational modification of the standard power law form for the primordial scalar spectrum, improves the fit to the data. Such improvements are much more significant w.r.t the standard modifications of the primordial spectra, obtained by considering a running spectral index. Let us note that the 2015 Planck data give constraints on the running, which are quite different from those coming from 2013 data. In particular the fit to the 2015 data does not improve much if one considers a running spectral index in the scalar sector.
On including the BK data in our analysis, we find that the results take vey similar values for the best fit. Furthermore  comparison with the data predicts, for our model, a loss in power of about $20-25\%$ w.r.t. the standard power law as $k$ approaches zero. and fixes the scale $\bar k$, which necessarily appears in the theoretical model. One finds  values for $\bar k$ which are very large, compared to the wave number associated with the largest observable scale in the CMB (namely $k_{\rm min}\simeq 1.4\cdot  10^{-4}\;{\rm Mpc}$). Let us note that the existence of such a small fundamental length may have relevant consequences on astrophysical observation. Indeed it is associated with distances which are comparable with the diameter of a large galaxy or a galaxy cluster. We further observe that a 3 order of magnitude variation of the value of $\bar k$ can be obtained on "re-tuning" the parameters used for its estimate. Further we observe that the value of $\bar k$, although illustrated for a specific inflationary model, is quite general and is found for diverse power loss compatible models. This is rather surprising and of course,assuming our proposed mechanism is correct, indicates  the possible presence of new physics at such scales. Actually such a result is not new. Indications for this have been seen both from a study of the stability of clusters of galaxies or is associated with the running of Newton's constant (\cite{scale}).\\

%%%%%%%%%%%%
\section*{Acknowledgments}
The work of A. K. was partially supported by the RFBR grant 14-02-00894.

%%%%%%%%%%%%%%%%%%%%%%%%%%%%

%%%%%%%%%%%%%%%%%%%%
%%%%%%%%%%%%%%%%%%%%
\end{document}